\documentclass[a4paper,11pt]{article}

\usepackage[T1]{fontenc}
\usepackage[utf8]{inputenc}
\usepackage{lmodern}
\usepackage{textcomp}

\usepackage{amsmath}
\usepackage{amssymb}
\usepackage{amsfonts}
\usepackage{bm}

\usepackage{graphicx}
\usepackage{float}
\usepackage[font=small,labelfont=bf]{caption}

\usepackage[numbers,sort&compress]{natbib}

\usepackage{authblk}

\usepackage[section]{placeins}

\usepackage{xcolor}

\usepackage[
  colorlinks=true,
  citecolor=blue,
  linkcolor=blue,
  urlcolor=blue,
  breaklinks=true
]{hyperref}

\usepackage{geometry}
\usepackage{indentfirst}

\newcommand{\ud}{\,\mathrm{d}}
\newcommand{\Reynolds}{\mathrm{Re}}
\newcommand{\Prandtl}{\mathrm{Pr}}
\newcommand{\Schmidt}{\mathrm{Sc}}
\newcommand{\Sherwood}{\mathrm{Sh}}

\begin{document}

\title{Evaporative Cooling and Secondary Combustion in a Liquid-Hydrocarbon Rotating Detonation Combustor with Dual-Purpose Head-End Injection}

\author[1,2]{Yongbin Zhu}
\author[1,2]{Songbai Yao\thanks{Corresponding author: yaosongbai@nimte.ac.cn (S. Yao)}}
\author[1,2]{Wenwu Zhang}
\affil[1]{Ningbo Institute of Materials Technology and Engineering, Chinese Academy of Sciences, Ningbo 315201, China}
\affil[2]{University of Chinese Academy of Sciences, Beijing 100049, China}

\date{}
\maketitle

\begin{abstract}
The rotating detonation engine (RDE) faces severe thermal protection
challenges. This
study numerically investigates a head-end dual-purpose injection configuration
in which the film-cooling holes act as both fuel and coolant injectors. An Eulerian--Lagrangian framework is employed, in which the
turbulence closure, the film-cooling effectiveness prediction, and the droplet
evaporation model are validated against experimental benchmarks. The validated framework is then applied to examine the effects
of coolant flow rate, atomization quality, and the fuel allocation between the
main flow and the cooling system. Beyond a threshold coolant flow rate the wall
temperature drops substantially, whereas an excessively low flow rate elevates
it by intensifying local deflagration, revealing a competition between evaporative cooling and secondary combustion that is inherent to this dual-purpose configuration. Fine droplets cool effectively but only over a limited distance
downstream, and fuel allocated to cooling weakens the detonation wave and
reduces thrust. A moderate coolant flow rate combined with fine atomization is therefore recommended.
\end{abstract}

\noindent\textbf{Keywords:} Rotating detonation; Liquid fuel; Secondary combustion; Film cooling; Thermal protection; Phase change; Coolant allocation; Cooling effectiveness

\section{Introduction}

Rotating detonation engines (RDEs) have emerged as a promising propulsion technology, offering higher thermodynamic efficiency and more compact engine configurations than conventional deflagration-based combustion systems~\cite{Wolanski2013,Raman2023}. Unlike traditional gas turbine combustors, where combustion occurs through subsonic flame propagation, RDEs use one or more continuously rotating detonation waves (RDWs) that travel circumferentially within an annular combustion chamber. The detonation process is characterized by a coupled shock wave and reaction zone, resulting in near-constant-volume heat release and pressure-gain combustion. These advantages render RDEs an attractive candidate for a broad range of propulsion and power-generation systems~\cite{IshiharaJSR2025a,TeasleyAS2F2023a,ZhangAeScTe2025a,PlaehnJPP2022a,PengApECS2025a,PalJEGTP2025a,WenCJA2024a,SmirnovAeScTe2018a,ZhouCoFl2023a,HanAeScTe2022a,JiaoPCI2024a}.

Despite considerable progress, several challenges remain before RDEs can be implemented in practice. Among them, the thermal environment within the combustor is particularly critical. Multiple experimental studies have consistently reported extreme heat flux levels: Falempin and Le Naour~\cite{Falempin2009} measured 12--17~MW/m$^{2}$ near the injection plane in a H$_{2}$/O$_{2}$ RDE; Theuerkauf et al.~\cite{Theuerkauf2014,Theuerkauf2016,Theuerkauf2015,Theuerkauf2013} reported peak values exceeding 9~MW/m$^{2}$; Stevens et al.~\cite{Stevens2018,Stevens2019} measured approximately 1.57~MW/m$^{2}$ on the outer wall of a hydrogen--air RDE; Azeman et al.~\cite{Azeman2023} found inner wall temperatures reaching 1100~$^{\circ}$C in a methane--oxygen RDE; and Micka et al.~\cite{Micka2021} recorded peak heat fluxes as high as 25~MW/m$^{2}$ near the injection plane in a high-pressure GOX--GH$_{2}$ rotating detonation rocket engine. These findings collectively underscore that effective thermal protection is a prerequisite for practical RDE deployment. Hou et al.~\cite{HouIJHMT2024a} and Lei et al.~\cite{LeiIJHE2025a} conducted unsteady conjugate heat transfer simulations that captured the instantaneous wall heat flux evolution of an RDE, and found that the peak heat flux differs between the inner and outer walls, with levels that could exceed 10~\text{MW/m}$^{2}$.

In response to this extreme environment, various active cooling strategies have been explored, among which film cooling, transpiration cooling, and regenerative cooling have shown considerable promise. For film cooling, researchers have examined both its cooling performance and its interactions with the detonation flow field. Tian et al.~\cite{Tian2022,Tian2026} showed that the detonation wavefront strongly blocks the coolant jet while the oblique shock waves deflect it periodically, and that an optimized annular-slit configuration can achieve a cooling effectiveness of 0.7--0.8. Yu et al.~\cite{Yu2023IJHE,Yu2024CatEar,Yu2023AST} demonstrated through combined numerical and experimental investigations that cat-ear-shaped holes outperformed cylindrical holes in both cooling effectiveness and protected area, establishing a consistent design-to-experiment validation framework. A particularly important aspect is the role of secondary combustion: Sridhara et al.~\cite{Sridhara2024} found that oxygen from the cooling air reacted in a deflagration with the unburned hydrogen left behind the detonation wave, raising the combustion efficiency from 90.5\% to 96.8\% while increasing the wall temperature by about 100~K. Zhou et al.~\cite{Zhou2026Energy} examined kerosene droplet mist cooling and found that partial secondary combustion did not negate the overall cooling performance, thereby demonstrating the feasibility of hydrocarbon-based film cooling. Additional studies have focused on geometric optimization. Lv et al.~\cite{Lv2024} showed that a longer liner mitigated the detonation-wave blocking effect and achieved a cooling effectiveness of 0.93. Shen et al.~\cite{Shen2024} found that a small injection angle of 30$^{\circ}$ provided optimal cooling performance and revealed an asymmetric cooling phenomenon between the inner and outer walls.

Transpiration cooling has also shown considerable potential in both experiments and simulations. Wen and Wang~\cite{Wen2020} showed that a porous wall with sufficient open-area ratio effectively suppressed acoustic modes and improved detonation stability. Frolov et al.~\cite{Frolov2020} showed that a porous ring feeding a liquid n-pentane film enabled continuous spinning detonation while providing active thermal protection through evaporative heat absorption. Lu et al.~\cite{Lu2025Phase,Lu2025Porous} further demonstrated that transpiration cooling coupled with phase-change heat transfer reduced wall temperatures from approximately 2500~K to below 500~K, achieving cooling effectiveness greater than 0.9, with water outperforming kerosene and optimal porosity depending on coolant type. Bian et al.~\cite{Bian2025} proposed a hybrid regenerative--transpiration cooling strategy and identified transpiration as the dominant heat absorption mechanism. Together, these studies confirm that transpiration cooling and its hybrid configurations offer significant advantages for RDE thermal protection. Regenerative cooling for rotating detonation engines has also advanced recently~\cite{SandriATE2TTCEBMU2023a}; for example, Jorgensen et al.~\cite{JorgensenAS2F2022a} introduced regenerative cooling into rotating detonation rocket engines and established a structural optimization framework that incorporates thermal stress, bending stress, and detonation-induced dynamic stress. Teasley et al.~\cite{TeasleyAS2F2023a} conducted hot-fire tests of a 7K lbf-thrust-class additively manufactured GRCop-42/84 copper alloy rotating detonation rocket engine at NASA Marshall Space Flight Center. Using both liquid oxygen/methane and liquid oxygen/hydrogen propellant combinations, they achieved a single burn lasting 133 seconds with deionized water and methane regenerative cooling, accumulating 802 seconds of total firing time across 18 starts.
Kawalec et al.~\cite{Kawalec2023} designed and tested a regenerative cooling jacket on a liquid-propellant N$_2$O/C$_3$H$_8$ rotating detonation engine, achieving 4 seconds of stable operation and a successful static firing, and ultimately powering the first rocket flight propelled entirely by a rotating detonation engine, as publicly reported.

Despite the progress outlined above, critical gaps remain in the thermal protection of liquid-hydrocarbon-fueled RDEs. In such engines, the coolant and fuel are typically the same liquid hydrocarbon, introducing a coupling between cooling and combustion that is absent when inert coolants (e.g., water or air) are used. On the one hand, the injected liquid hydrocarbon can absorb substantial heat through phase change and wall film coverage, providing effective thermal protection. On the other hand, the heated coolant may undergo secondary combustion with the unburned species behind the detonation wave, releasing additional heat near the wall and partially offsetting the cooling benefit. Although this competing influence has been observed for air film cooling~\cite{Sridhara2024} and kerosene droplet cooling~\cite{Zhou2026Energy}, the underlying mechanisms remain unresolved. Specifically, the spatial and temporal coupling between phase-change heat absorption and secondary combustion heat release, and its dependence on local detonation wave structure, have not been systematically characterized. Furthermore, design criteria for liquid hydrocarbon jets under highly unsteady detonation waves have not been established; it remains unclear how jet trajectory, breakup, and film formation respond to the periodic passage of the detonation and oblique shock waves, or how these responses govern the cooling and combustion behavior.

To address these gaps, the present study numerically investigates the thermal protection performance of a liquid-hydrocarbon-fueled rotating detonation combustor (RDC) featuring a novel head-end dual-purpose configuration. In this design, liquid film-cooling holes are placed exclusively in the head region of the combustor, which is the primary path swept by the RDW and therefore the zone of the most intense detonation activity. These injection sites serve a dual purpose, acting as both coolant and fuel injectors, because the liquid hydrocarbon is supplied through the film-cooling holes. Once injected, the liquid fuel forms a near-wall film that provides thermal protection through evaporative cooling and wall coverage, while also participating directly in the detonation-driven combustion process. In particular, the injected fuel is exposed to the highly unsteady detonation and post-detonation flow field, where it can undergo secondary combustion with the residual oxidizer or the unburned species. Characterizing this secondary combustion and its two competing effects on wall heat transfer, namely cooling by endothermic phase change and heating by exothermic reaction, is a central objective of the present work. 

Overall, the objectives of the present study are twofold. The first is to characterize the coupled effects of coolant flow rate, atomization quality, and fuel allocation on the thermal and propulsion performance of this dual-purpose injection configuration. The second is to establish design criteria that balance cooling effectiveness with combustion efficiency under unsteady rotating detonation wave conditions. Through a systematic parametric study, the present work aims to provide physical insight into liquid-film-cooled RDEs.
\section{Numerical framework and validation}

\subsection{RDC configuration and operating conditions}

The combustor has a coaxial annular configuration, an inner radius of 30~mm, an outer radius of 36~mm, and an axial length of 60~mm. The annular surface at the
bottom of the combustor serves as the main inlet, through which the majority
of the fuel and all of the oxidizer are supplied by full-face injection. The oxidizer is oxygen-enriched air with an oxygen mass fraction of
0.5, injected at a total pressure of 8~atm and a total temperature of 300~K.
The fuel is liquid kerosene, injected as droplets normal to the bottom surface at 50~m/s and 300~K, matching the oxidizer temperature. To reduce the thermal load on the outer wall,
four layers of coolant injection holes are arranged along the axial direction
on the outer wall of the combustor. Each layer consists of 36 uniformly
distributed injection holes with a radius of 0.3~mm. The first layer of holes
is located at $z = 2$~mm, with an axial spacing of 3~mm between adjacent
layers. All injection holes are positioned below the detonation wave head
height. The
annular surface at the top of the combustor serves as the main outlet, where the high-temperature, high-pressure combustion products are discharged into the ambient environment at 1~atm and 300~K.

\begin{figure}[!htbp]
\centering
\includegraphics[width=0.5\textwidth]{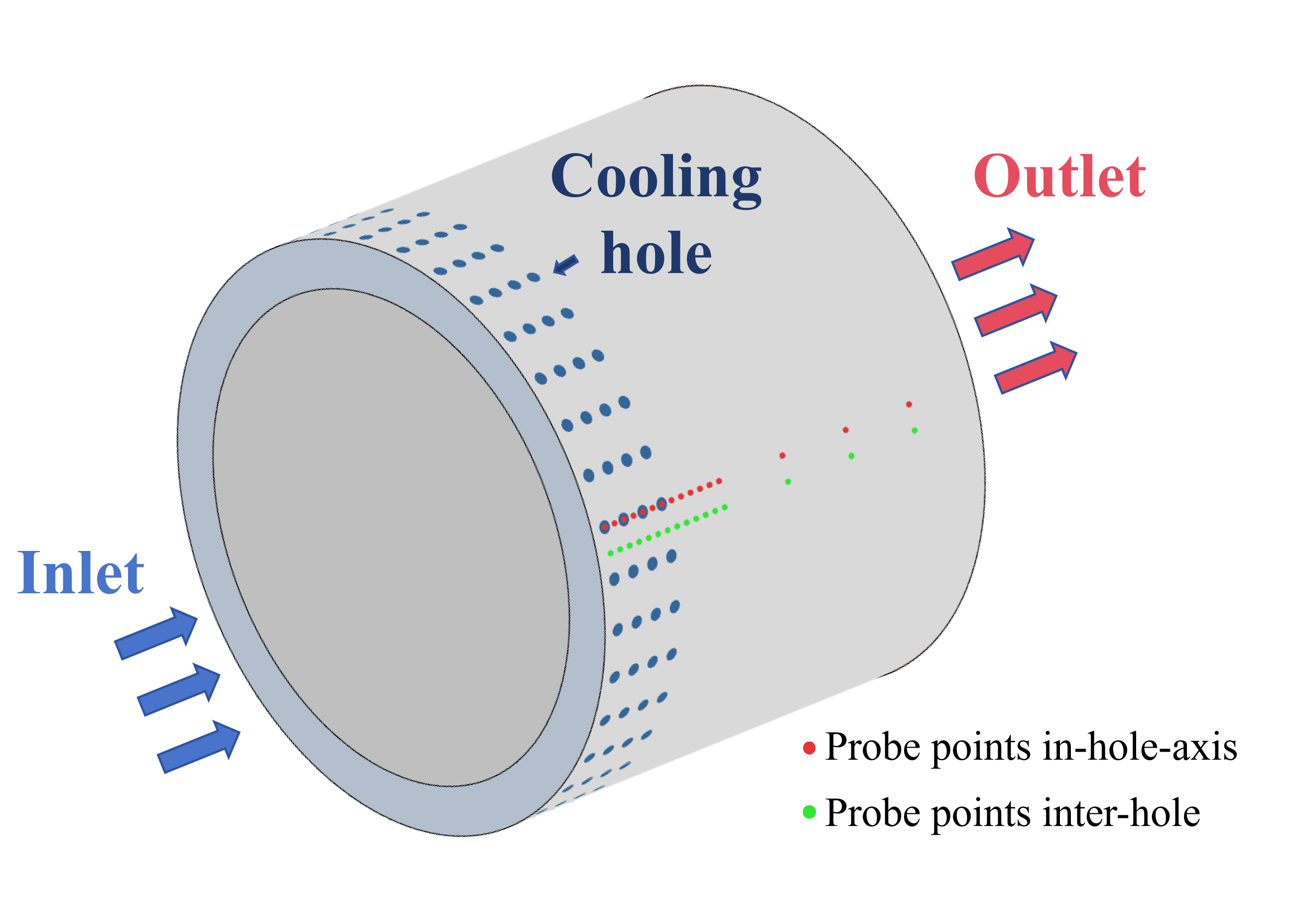}
\caption{Schematic of the RDC with head-end film-cooling holes and wall monitoring points}
\label{fig:combustor}
\end{figure}

To monitor the wall conditions, two rows of monitoring points are arranged along the axial direction on the outer wall. One row (red) lies on the same axial lines as the cooling holes, i.e., aligned with the center of a cooling hole. The other row (green) lies on the axial lines midway between adjacent cooling holes, i.e., shifted circumferentially by half a hole spacing. Both rows share the same axial coordinates, so that the in-hole axis and inter-hole locations can be compared directly. The overall layout of the RDC, including the head-end film-cooling holes and the two rows of wall monitoring points, is shown in Figure~\ref{fig:combustor}.

\subsection{Governing equations and numerical models}
\label{sec:gov_eq}

The simulations are performed with a custom finite-volume solver built on the OpenFOAM library and derived from rhoCentralFoam. The flow is solved in a Reynolds-averaged Navier--Stokes (RANS) framework,
and turbulence is closed with the $k$--$\omega$ shear-stress transport (SST)
model of Menter~\cite{MenterAIAAJ1994a}.
The Kurganov--Tadmor second-order central-upwind scheme is used for flux evaluation, and time integration is performed with the second-order implicit backward scheme. The species transport equations
are solved together with the flow equations, and the stiff chemical source terms are integrated with the Euler implicit scheme. This OpenFOAM-based numerical framework is widely applied
for high-fidelity simulations of flow fields in RDEs.
The governing equations for the gas phase are as follows:
\begin{equation}
\frac{\partial\rho}{\partial t} + \nabla \cdot (\rho\mathbf{U}) = S_{m}
\label{eq:continuity}
\end{equation}
\begin{equation}
\frac{\partial}{\partial t}(\rho\mathbf{U}) + \nabla \cdot (\rho\mathbf{UU}) + \nabla p = \nabla \cdot \tau_{eff} + S_{u}
\label{eq:momentum}
\end{equation}
\begin{equation}
\frac{\partial}{\partial t}(\rho E) + \nabla \cdot [\,\mathbf{U}(\rho E + p)\,] = \nabla \cdot (\tau_{eff} \cdot \mathbf{U}) - \nabla \cdot \mathbf{q} + S_{h} + S_{e}
\label{eq:energy}
\end{equation}
\begin{equation}
\frac{\partial}{\partial t}(\rho Y_{n}) + \nabla \cdot (\rho Y_{n}\mathbf{U}) + \nabla \cdot (\mathbf{s}_{n}) = \dot{\omega}_{n} + S_{Y}
\label{eq:species}
\end{equation}
In Eqs.~\eqref{eq:continuity}--\eqref{eq:species}, $\rho$ is the density, $p$ the pressure, $E$ the total energy, and $\mathbf{U}$ the velocity vector. The effective viscous stress tensor is given by
\[
\tau_{eff} = \mu_{eff}\bigl[ (\nabla\mathbf{U}) + (\nabla\mathbf{U})^{\mathrm{T}} - \tfrac{2}{3}(\nabla\cdot\mathbf{U})\mathbf{I} \bigr],
\]
where the effective viscosity $\mu_{eff} = \mu + \mu_{t}$. Under Sutherland's
law, the laminar viscosity $\mu$ is computed as
\begin{equation}
    \mu = \frac{A_{s}\sqrt{T}}{1 + T_{s}/T},
\label{eq:sutherland}
\end{equation}
with coefficients $A_{s} = 1.67212\times10^{-6}~\mathrm{kg/(m{\cdot}s{\cdot}\sqrt{K})}$
and $T_{s} = 170.672~\mathrm{K}$. The turbulent viscosity is obtained from the
$k$--$\omega$ SST model of Menter~\cite{MenterAIAAJ1994a}, and the heat flux follows Fourier's law, $\mathbf{q} = -k_{eff}\nabla T$, where the
effective thermal conductivity $k_{eff}$ is evaluated using the Eucken
approximation for the laminar part, and the turbulent contribution is $k_{t} = \mu_{t}C_{p}/\Prandtl_{t}$. $\mathbf{s}_{n} = -D\nabla(\rho Y_{n})$ is the mass flux
vector with the mass diffusivity computed as $D = k/\rho C_{p}$. $S_{h}$ is the heat release rate of the chemical reactions, $Y_{n}$ the mass fraction of species $n$ ($n = 1,2,\dots,N_{S}$, with $N_{S}$ the number of species), and $\dot{\omega}_{n}$ the net production rate of species $n$.

The total energy is defined as
\begin{equation}
E = \sum_{i=1}^{N_{S}} Y_{i}h_{i} - \frac{p}{\rho} + \frac{1}{2}(u^{2}+v^{2}+w^{2})
\label{eq:totalE}
\end{equation}
\begin{equation}
h_{i} = h_{f,i}^{0} + \int_{T_{\mathrm{ref}}}^{T} C_{p,i}(T)\,\ud T
\label{eq:enthalpy}
\end{equation}
$T_{\mathrm{ref}}$ and $h_{f,i}^{0}$ denote the reference temperature and the
standard-state enthalpy, respectively; $h_{i}$ is the specific enthalpy of
the $i$th species; and $C_{p,i}$ the specific heat capacity at constant pressure of the $i$th species.

The droplets are described by the discrete phase model (DPM) and obey the following Lagrangian equations:
\begin{equation}
\frac{dm_{d}}{dt} = -\dot{m}_{d}
\label{eq:droplet_mass}
\end{equation}
\begin{equation}
\frac{d\mathbf{u}_{d}}{dt} = \frac{\mathbf{F}_{d}}{m_{d}}
\label{eq:droplet_momentum}
\end{equation}
\begin{equation}
C_{p,d}\,\frac{dT_{d}}{dt} = \frac{\dot{Q}_{d} - \dot{m}_{d}\,h(T_{d})}{m_{d}}
\label{eq:droplet_energy}
\end{equation}
where $\mathbf{u}_{d}$ is the droplet velocity vector, $m_{d}$ is the droplet
mass, $C_{p,d}$ is the droplet heat capacity, and $h(T_{d})$ the fuel vapor enthalpy evaluated at the droplet surface. The evaporation mass transfer rate is $\dot{m}_{d}$. According to the film theory of
droplet evaporation~\cite{SpaldingSC481953a}, $\dot{m}_{d}$ is given by
\begin{equation}
\dot{m}_{d} = \pi D_{d}\,\Sherwood\,D_{\mathrm{vap}}\,\rho_{f}\ln(1+B_{m})
\label{eq:evap}
\end{equation}
where
$B_{m} = (X_{\mathrm{vap},s} - X_{\mathrm{vap},g})/(1 - X_{\mathrm{vap},s})$
is the Spalding mass transfer number, and $X_{\mathrm{vap},s}$ and
$X_{\mathrm{vap},g}$ are the mole fractions of fuel vapor on the droplet
surface and in the ambient gas, respectively. According to Raoult's law,
$X_{\mathrm{vap},s} = X_{d,i}\,p_{\mathrm{sat}}/p$, where $X_{d,i}$ is the
mole fraction of species $i$ in the liquid droplet and $p_{\mathrm{sat}}$ is
the saturated vapor pressure. $D_{\mathrm{vap}}$ is the vapor mass
diffusivity, $\rho_{f}$ is the density of the film around the droplet. The
Sherwood number is given by
\begin{equation}
\Sherwood = 2.0 + 0.6\,\Reynolds_{d}^{1/2}\,\Schmidt^{1/3}
\label{eq:Sherwood}
\end{equation}
The ideal gas law is used to evaluate the vapor density at the droplet
surface, and the present implementation also includes a boiling model
based on Ref.~\cite{ZuoIJER2000a} for regimes where the droplet temperature
approaches the boiling point. These models are validated in Section~\ref{sec:evap_model}.

Several forces act on a droplet. For the spherical droplets assumed here, the drag is the dominant contribution, given by
\begin{equation}
\mathbf{F}_{d} = \frac{18\mu}{\rho_{d}D_{d}^{2}}\,\frac{C_{d}\,\Reynolds_{d}}{24}\,m_{d}\,|\mathbf{u} - \mathbf{u}_{d}|
\label{eq:drag}
\end{equation}
where $\rho_{d}$ and $D_{d}$ are the droplet density and diameter,
respectively, and the droplet Reynolds number is
\begin{equation}
\Reynolds_{d} = \frac{\rho D_{d}\,|\mathbf{u} - \mathbf{u}_{d}|}{\mu}
\label{eq:Re_d}
\end{equation}
The drag coefficient $C_{d}$ is computed as
\begin{equation}
C_{d} = 
\begin{cases}
0.424, & \Reynolds_{d} > 1000 \\[4pt]
\displaystyle\frac{24}{\Reynolds_{d}}\Bigl(1 + \frac{1}{6}\,\Reynolds_{d}^{2/3}\Bigr), & \Reynolds_{d} \leq 1000
\end{cases}
\label{eq:Cd}
\end{equation}
$\dot{Q}_{d}$ is the convective heat transfer rate, given by
\begin{equation}
\dot{Q}_{d} = h_{c}A_{d}(T - T_{d})
\label{eq:Qdot}
\end{equation}
where $A_{d}$ represents the droplet surface area, $T$ is the temperature of
the continuous phase, $T_{d}$ is the droplet temperature, and $h_{c}$ is the
convective heat transfer coefficient calculated using the Ranz--Marshall
correlation~\cite{RanzCEP1952a,RanzCEP1952b}.
$S_{m}$, $S_{u}$, $S_{e}$, and $S_{Y}$ are the source terms of the discrete phase, accounting for the exchange of mass, momentum, energy, and species between the dispersed liquid phase and the carrier gas:
\begin{equation}
S_{m} = \frac{1}{V_{\mathrm{cell}}}\sum_{d} \dot{m}_{d}
\label{eq:Sm}
\end{equation}
\begin{equation}
S_{u} = -\frac{1}{V_{\mathrm{cell}}}\sum_{d} \mathbf{F}_{d}
\label{eq:Su}
\end{equation}
\begin{equation}
S_{e} = -\frac{1}{V_{\mathrm{cell}}}\sum_{d} \bigl( \dot{Q}_{d} - \dot{m}_{d}\,h(T_{d}) \bigr)
\label{eq:Se}
\end{equation}
\begin{equation}
S_{Y} = 
\begin{cases}
S_{m}, & \text{for kerosene species} \\
0,    & \text{for other species}
\end{cases}
\label{eq:SY}
\end{equation}
where $V_{\mathrm{cell}}$ is the control volume.
The chemical reaction between kerosene and air is modeled using the two-step mechanism of Franzelli et al.~\cite{Franzelli2010}:
\begin{align}
\mathrm{KERO} + 10\,\mathrm{O}_{2} &\rightarrow 10\,\mathrm{CO} + 10\,\mathrm{H}_{2}\mathrm{O} \label{eq:rxn1} \\
\mathrm{CO} + 0.5\,\mathrm{O}_{2} &\rightleftharpoons \mathrm{CO}_{2} \label{eq:rxn2}
\end{align}
The reaction rates are computed using the model parameters from
Ref.~\cite{Franzelli2010}. This two-step mechanism has been used widely in previous studies, including detonation simulations.

\FloatBarrier
\subsection{\texorpdfstring{Grid independence analysis}{Grid independence analysis}}
\begin{figure}[!htbp]
\centering
\includegraphics[width=0.5\textwidth]{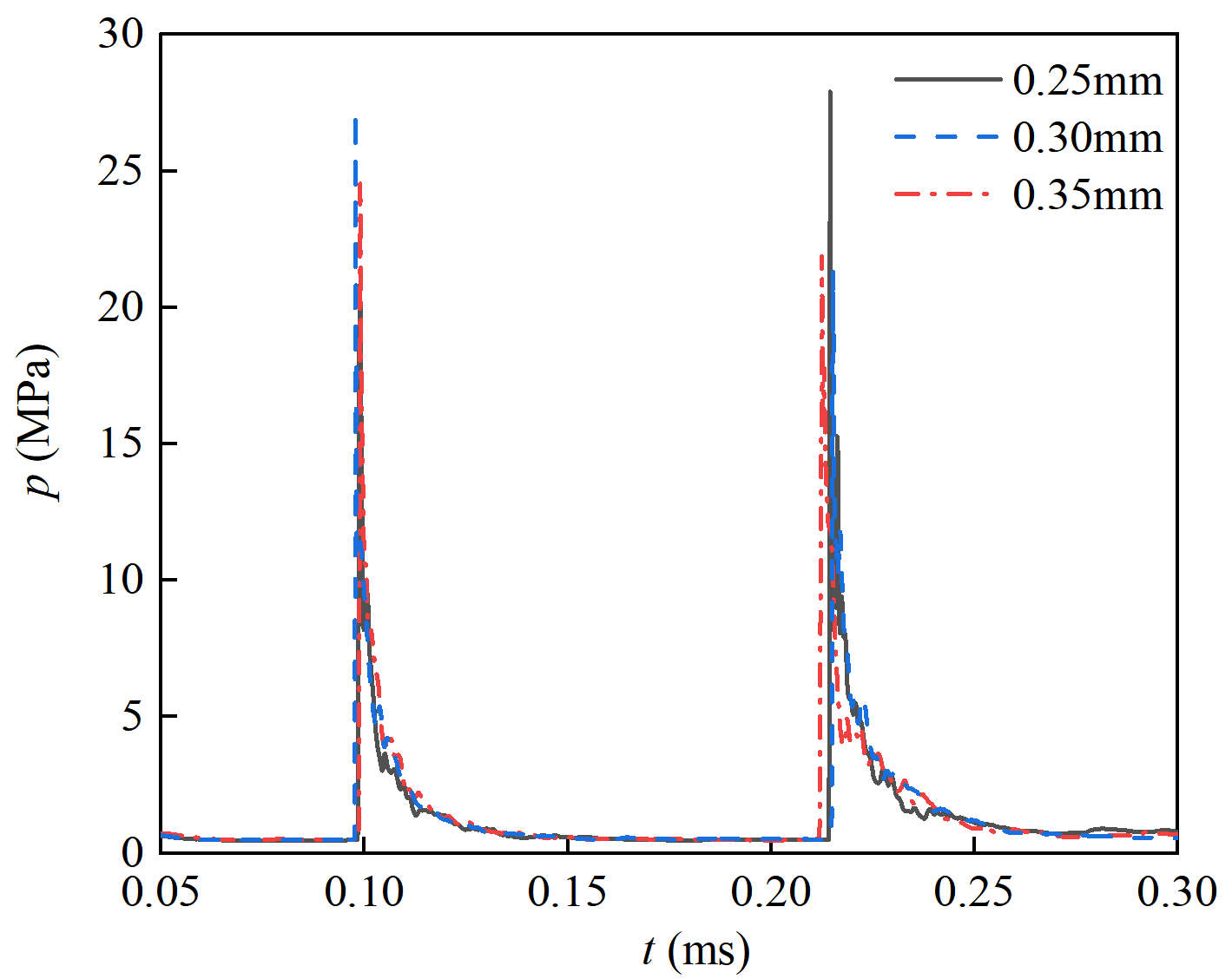}
\caption{Grid independence study: pressure profiles of rotating detonation waves for three mesh resolutions}
\label{fig:grid_pressure}
\end{figure}

\begin{figure}[!htbp]
\centering
\includegraphics[width=4.8in,height=2.874in]{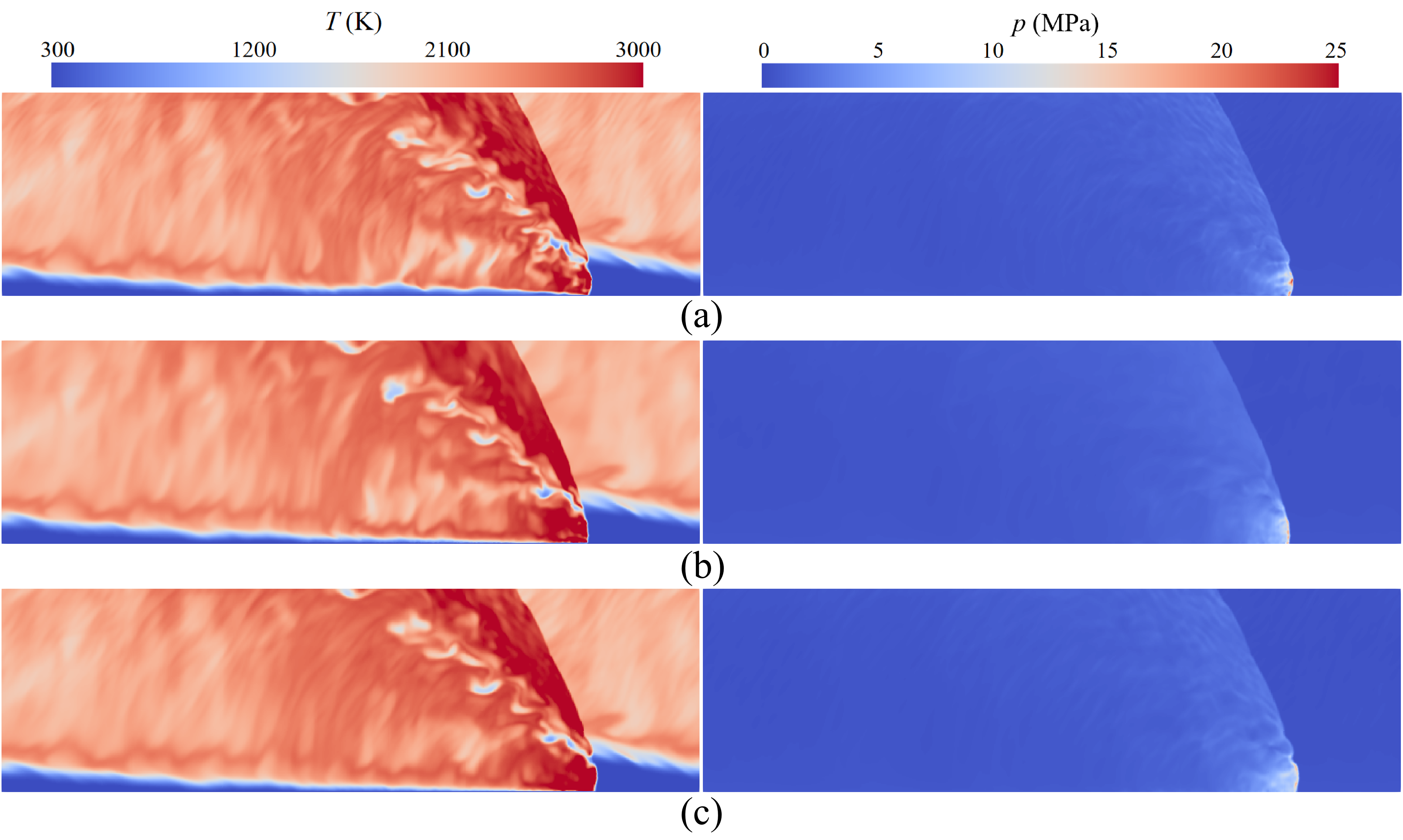}
\caption{Grid independence study: temperature and pressure distributions for three mesh resolutions:
(a) $\Delta x = 0.25$~mm, (b) $\Delta x = 0.3$~mm, and (c) $\Delta x = 0.35$~mm}
\label{fig:grid_contour}
\end{figure}

\label{sec:grid}
Grid independence was examined with three meshes whose maximum cell sizes are 0.25, 0.3, and 0.35~mm, using identical boundary conditions and initial fields.
Figure~\ref{fig:grid_pressure} presents the temporal evolution of the pressure at $R=35$~mm and $z=5$~mm. The resulting detonation velocities are 1900.85, 1871.20, and 1934.32~m/s, with a maximum relative deviation below 5\% among the three meshes. Figure~\ref{fig:grid_contour} shows the temperature and pressure contours unwrapped on the cylindrical surface at $R=33$~mm. All three meshes resolve the main flow structures, including the detonation wavefront, the oblique shock waves, and the slip lines. To balance accuracy against cost, a maximum cell size of 0.3~mm is adopted for all the simulations reported below.

\FloatBarrier
\subsection{Turbulence and film-cooling model validation}
\label{sec:turb_validation}

The reliability of the numerical framework adopted here is assessed against benchmarks. The classical backward-facing step flow of Vogel \& Eaton~\cite{Vogel1985} is
adopted as the test case. It is selected because it embodies the canonical flow
physics of separation, recirculation, and reattachment, and because the wall
shear stress is measured directly in the same experiment. These features are
highly relevant to the near-wall flow in the head-end region of the RDE, where
the injected liquid coolant interacts with the main crossflow and may induce
localized separation. An accurate prediction of the wall shear stress
distribution in such separated flows is critical here, as it directly governs
film spreading, droplet attachment, and the thermal resistance in the near-wall
region.

The experiment of Vogel \& Eaton is one of the few benchmark-quality data sets
that provide detailed measurements of the mean flow, the turbulence statistics,
and the wall heat transfer for the same separated flow. The measurements were
obtained in a channel with an expansion ratio of 1.25, defined as the ratio of
the downstream to the upstream channel height, so that the upstream and
downstream channel heights are $4H$ and $5H$, respectively, where $H$ denotes
the step height. The Reynolds number based on the step height and the upstream
reference velocity is $\Reynolds_{H} = 28\,000$, the working fluid is air with
$\Prandtl = 0.71$, and a uniform heat flux of 270~W/m$^{2}$ is imposed on the
wall downstream of the step. Because of its fully documented geometry,
boundary conditions, and uncertainty, this case has been adopted as a standard
validation case for RANS closures in independent studies with different solvers and model families~\cite{JardineUKHTC2021}. The computational domain and boundary conditions are set to reproduce the experimental configuration.

\begin{figure}[!htbp]
\centering
\begin{minipage}[b]{0.53\textwidth}
\centering
\includegraphics[width=\linewidth]{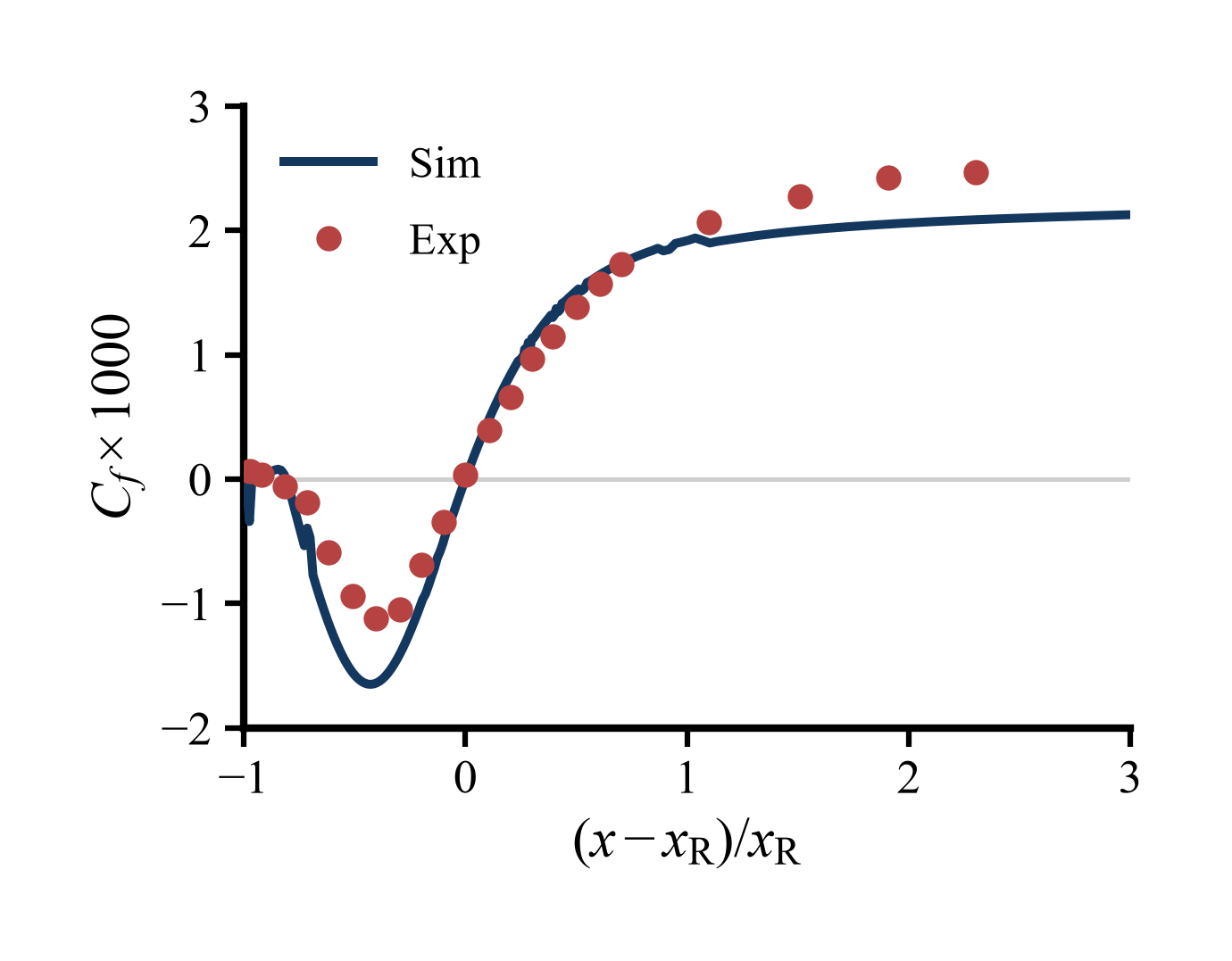}\\[1pt]
{\small\bfseries (a)}
\end{minipage}\hfill
\begin{minipage}[b]{0.42\textwidth}
\centering
\includegraphics[width=\linewidth]{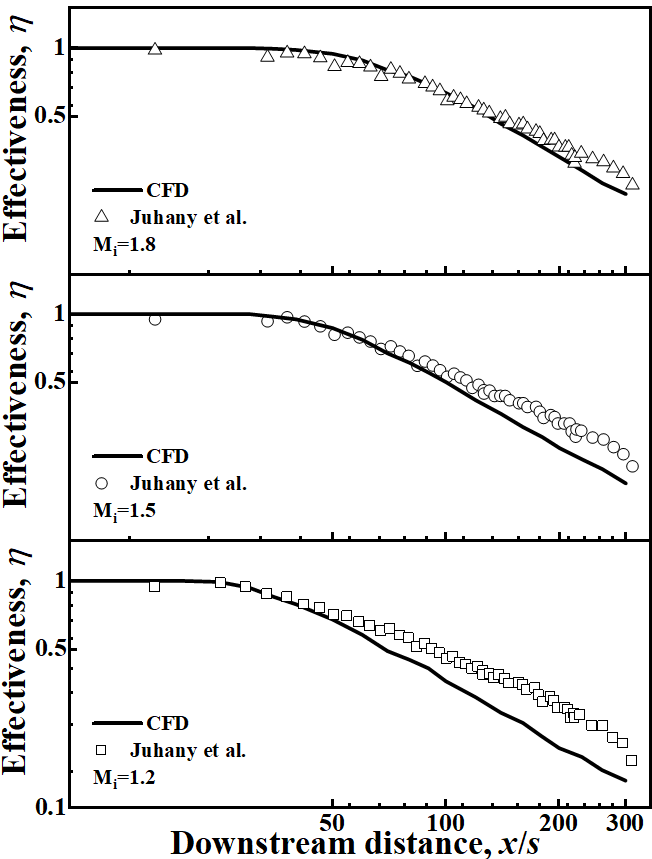}\\[1pt]
{\small\bfseries (b)}
\end{minipage}
\caption{(a) Comparison of the predicted time-averaged skin-friction coefficient $C_{f}$ with the experimental data of Vogel \& Eaton~\cite{Vogel1985} for the backward-facing step flow, and (b) comparison of the predicted film-cooling effectiveness~\cite{LiArXiv2026} with the experimental data of Juhany et al.~\cite{Juhany1994} for supersonic film cooling at injectant Mach numbers $M_{i}=1.2$, 1.5, and 1.8}
\label{fig:turbulence}
\end{figure}

Figure~\ref{fig:turbulence}(a) presents the distribution of the time-averaged skin-friction coefficient $C_{f}$ along the streamwise direction, together with
the experimental data of Vogel \& Eaton for comparison, where
$C_{f} = \tau_{w}/(0.5\rho_{\infty}U_{\mathrm{ref}}^{2})$ and $\tau_{w}$ is the
wall shear stress. The predicted $C_{f}$ distribution agrees well with the
measurements. In particular, the numerical method successfully captures the
negative skin-friction coefficient within the recirculation zone, the peak
negative $C_{f}$ being located at $(x - x_{R})/x_{R} = -0.4$, in good agreement
with the experimental data. The time-averaged reattachment point, defined as the location where $C_{f}=0$, is predicted to be at $x_{R}/H = 6.35$, which
deviates by less than 5\% from the experimental value of $x_{R}/H = 6.67$.

This deviation is of the same order as that reported for other closures
applied to the same case~\cite{JardineUKHTC2021}, which suggests that the
residual discrepancy is characteristic of the closure family employed here
rather than of the present implementation. However, this comparison addresses the momentum side of the closure through $C_{f}$, and the backward-facing step case is a low-speed, incompressible, single-phase flow, whereas the head-end flow of the present RDC is compressible, multiphase, and strongly unsteady. The comparison thus
establishes that the closure reproduces separation, reattachment, and the
associated wall shear stress in a canonical separated flow; it is not intended
to establish similarity with the RDC conditions themselves. Compressibility
effects are instead addressed by the film-cooling validation below, and the
two-phase behavior by the droplet evaporation validation in the following
subsection.

The film-cooling treatment adopted here is assessed against the supersonic air-film-cooling experiments of Juhany et al.~\cite{Juhany1994}, shown in Figure~\ref{fig:turbulence}(b). These experiments
correspond to a freestream Mach number of $M_{\infty}=2.44$, a coolant-slot
height of $s=1.5$~mm, and injectant Mach numbers of $M_{i}=1.2$, 1.5, and 1.8,
with the coolant static pressure matched to that of the freestream, and they
provide well-documented measurements of the film-cooling effectiveness on a
compressible wall. This validation was carried out in the authors' previous
study of film-cooled oblique detonation combustors~\cite{LiArXiv2026}, which
employed the same solver; the results are
reproduced here for completeness. As shown, the predicted film-cooling
effectiveness reproduces the streamwise decay of the measurements and its
dependence on the injectant Mach number, with mean absolute deviations of 0.087, 0.057, and 0.035 for $M_{i}=1.2$, 1.5, and 1.8, respectively, so that the deviation decreases as the injectant Mach number increases. These results
support the use of the present near-wall treatment for the film-cooling
simulations discussed below.

Taken together with the grid independence study presented above, these two
benchmarks bound the modeling uncertainty of the present framework: the grid
study constrains the discretization error, the backward-facing step case
verifies the prediction of the near-wall flow in a separated region, and the
supersonic film-cooling case verifies the prediction of the wall thermal
response under compressible conditions.

\FloatBarrier
\subsection{Droplet evaporation model validation}
\label{sec:evap_model}

The droplet evaporation model is validated against measurements of single n-heptane droplet evaporation in microgravity. The conditions follow the experiments of Nomura et al.~\cite{NomuraSC211996a}, with an ambient pressure of 0.1~MPa and ambient temperatures of 471, 555, and 648~K.
These temperatures span the critical temperature of n-heptane ($T_{\mathrm{cr}} = 540$~K).

\begin{figure}[!htbp]
\centering
\includegraphics[width=0.5\textwidth]{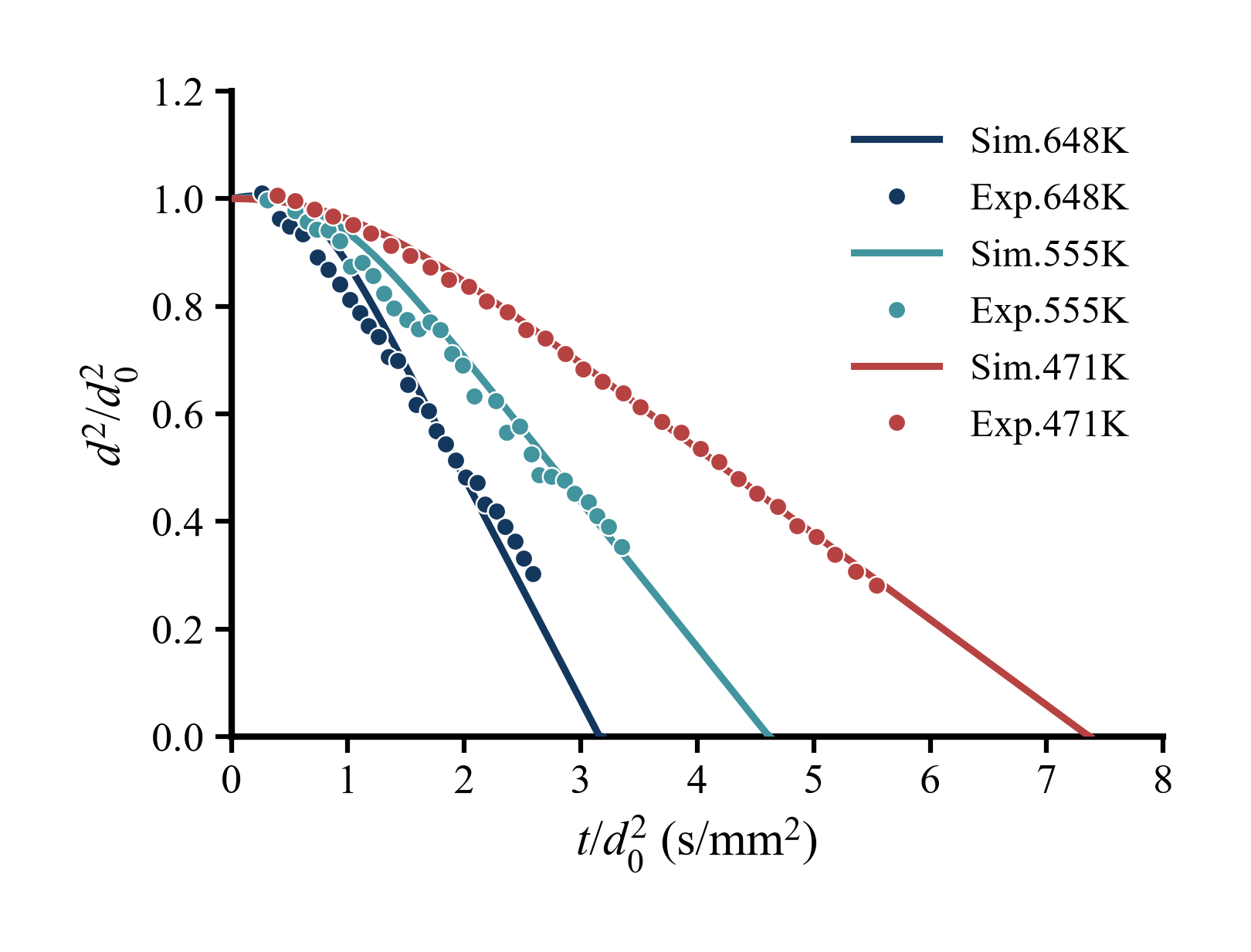}
\caption{Evolution of the normalized squared droplet diameter $d^{2}/d_{0}^{2}$ as a function of normalized time $t/d_{0}^{2}$}
\label{fig:evap_model}
\end{figure}

Figure~\ref{fig:evap_model} presents the evolution of the normalized squared
droplet diameter $d^{2}/d_{0}^{2}$ (here $d$ and $d_0$ respectively denote
the instantaneous and initial droplet diameters, i.e.\ $D_d$ and $D_{d,0}$
in the notation of Section~\ref{sec:gov_eq}) as a function of the normalized time
$t/d_{0}^{2}$ under different ambient temperatures. The predictions agree well with the experimental data over the entire time range. For all three temperatures, the predicted $d^{2}/d_{0}^{2}$ curves decrease linearly, as expected from the $d^{2}$ law, and their slopes show excellent agreement with the measured values. At the higher temperature of 648~K the evaporation rate is significantly higher, which the steeper predicted slope reproduces.

These results indicate that the evaporation model reproduces the measured evaporation history over the temperature range examined and provide the basis for the spray-combustion simulations reported below.

\section{Results and discussion}

\begin{figure}[!htbp]
\centering
\includegraphics[width=4.17679in,height=2.46957in]{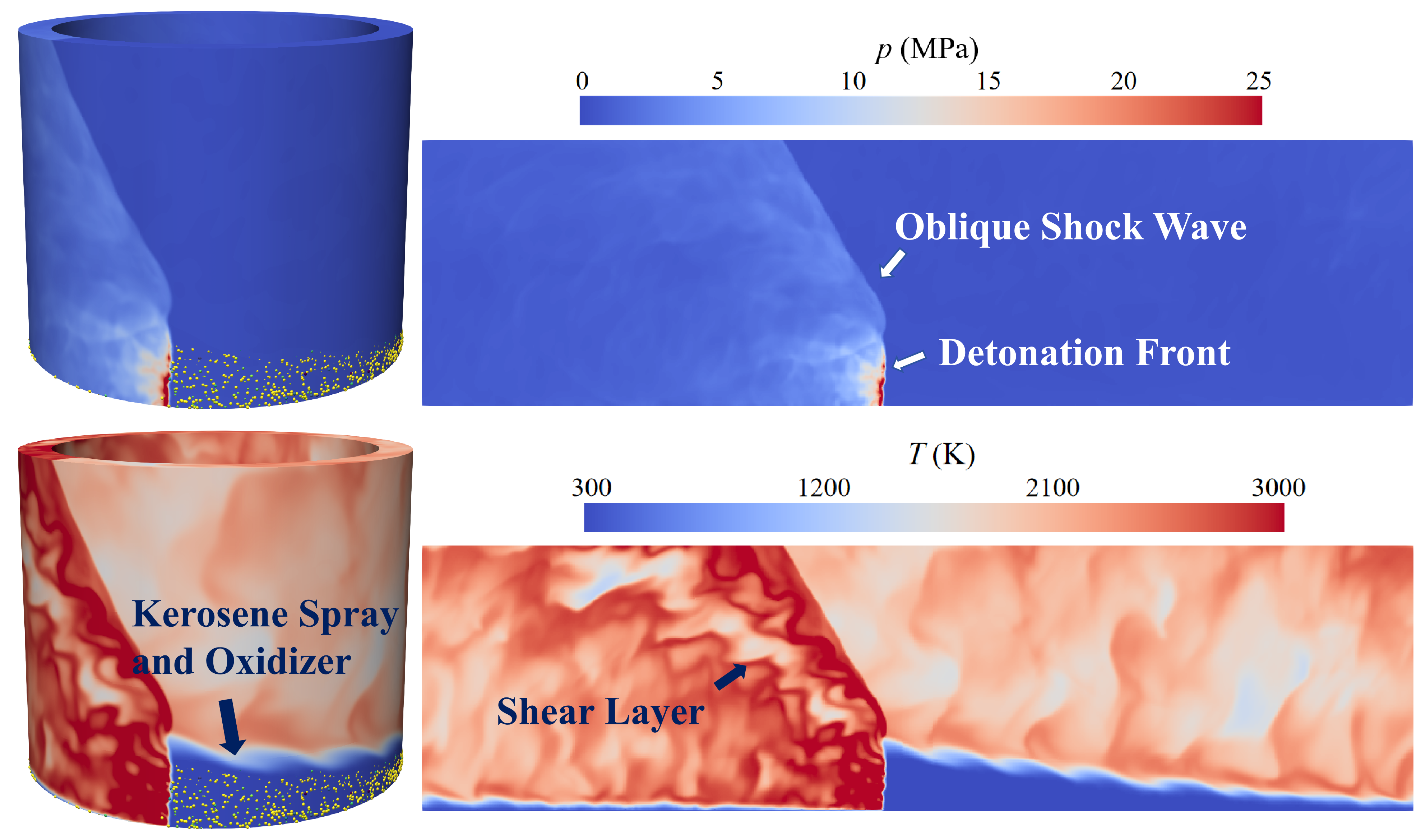}
\caption{Schematic of the RDW structure in the head-film-cooled RDC}
\label{fig:detonation_wave}
\end{figure}

A schematic of the detonation wave structure is shown in
Figure~\ref{fig:detonation_wave}. Most of the fuel (kerosene) is injected as droplets together with the oxidizer from the bottom surface of the combustor, and combustion occurs as the detonation wave
sweeps over the mixture. Since both the liquid droplets and the oxidizer have
an initial temperature of 300~K, the vapor concentration of kerosene in the
fresh mixture ahead of the detonation wave is extremely low, with virtually no
pre-evaporation. As the detonation wave passes, the droplets evaporate rapidly
and subsequently react with the oxidizer to complete combustion. Consequently,
the wavefront in two-phase RDEs is less smooth
than that in gaseous RDEs, exhibiting irregular features. Nevertheless, as
shown in Figure~\ref{fig:detonation_wave}, typical flow structures such as the
detonation wave, oblique shock waves, and shear layer can still be clearly
identified. Due to significant three-dimensional effects, radial disturbances
are non-negligible. Under such conditions, vortex structures induced by the
Kelvin--Helmholtz instability near the slip line can be observed, albeit with
poor morphological regularity. Additionally, a local low-temperature region
exists in the vicinity of the shear layer. This occurs because the oxidizer reaches a slightly greater height within the fresh mixture layer than the droplets do, so that the oxidizer at the top of the layer lacks sufficient fuel for combustion and forms a low-temperature, incompletely burned zone.

In Figure~\ref{fig:detonation_wave}, ``Kerosene Spray and Oxidizer'' denotes
the mixture of fuel droplets and oxidizer entering the flow field from the
bottom of the combustor. The parameters of the
droplets injected with the main flow at the bottom are not the primary focus here. To obtain a more stable two-phase detonation flow field, the diameter of these droplets is uniformly set to 5~$\mu$m, since smaller droplets mitigate the unsteadiness of the detonation wave. In contrast, the kerosene coolant droplets injected from the
wall are the main focus of this investigation. For this coolant, the effects of parameters such as the coolant flow rate, atomization quality, spatial distribution, and fuel flow distribution on the cooling effectiveness are analyzed systematically.

The wall temperature, cooling effectiveness, thrust, and species concentrations reported below are obtained by averaging the data from the monitoring points over multiple detonation cycles, which reduces the influence of the inherent transient instabilities of the detonation wave. Specifically, the wall
temperature is obtained by averaging the temperatures recorded at wall
monitoring points over multiple cycles; thrust and species concentration data
are obtained by averaging the measurements at the outlet monitoring surface
over multiple cycles. The cooling effectiveness $\eta$ and the thrust $F$ are defined as follows:

\begin{equation}
\eta = \frac{T_{w,0} - T_{w,c}}{T_{w,0} - T_{c}}
\label{eq:cooling_efficiency}
\end{equation}

\begin{equation}
F = \int_{A_{\mathrm{out}}} \bigl( \rho v_{z}^{2} + p - p_{\infty} \bigr) \,\ud A
\label{eq:thrust}
\end{equation}
where $T_{w,0}$ is the wall temperature in the no-cooling case,
$T_{w,c} = (T_{w,h1}+T_{w,h2})/2$ is the average wall temperature under cooled
conditions (with $T_{w,h1}$ and $T_{w,h2}$ being the temperatures at the in-hole axis and inter-hole locations, respectively), $T_{c} = 300$~K is the coolant
temperature, $\rho$ is the density of the exhaust gas, $v_{z}$ is the axial
velocity at the outlet, $p$ is the static pressure at the outlet,
$p_{\infty}$ is the ambient pressure, and $A_{\mathrm{out}}$ is the outlet
area.

\FloatBarrier
\subsection{Effect of coolant flow rate}
\label{sec:flowrate}

The coolant flow rate is one of the key factors affecting the cooling effectiveness.
To investigate this, numerical simulations are conducted for various coolant
flow rates while keeping the fuel flow rate injected with the oxidizer in the
main flow constant at 0.15~kg/s. The coolant flow rate is varied from 0 to
0.10~kg/s. The coolant is uniformly distributed among 144 injection holes
arranged in four layers ($4\times36$), with a coolant droplet diameter of
10~$\mu$m.

\begin{figure}[!htbp]
\centering
\includegraphics[width=5.808in,height=2.31349in]{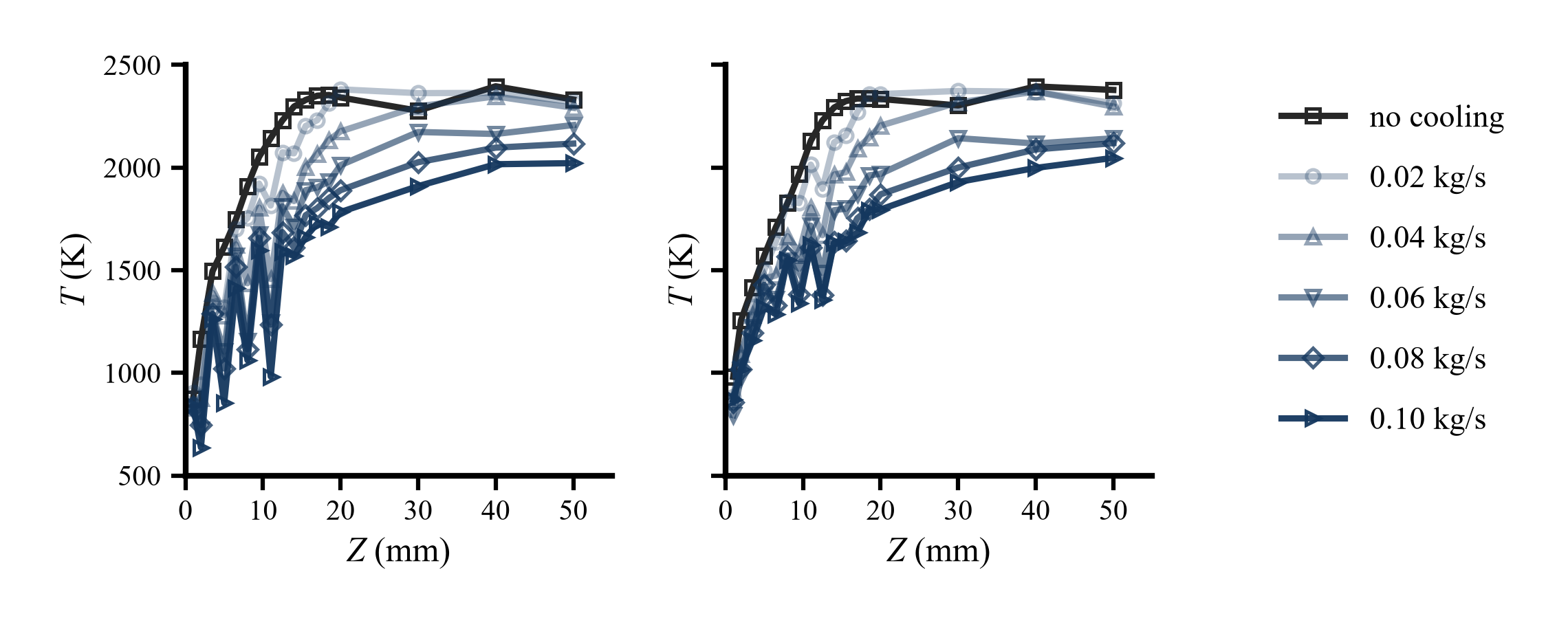}
\caption{Wall temperature distributions at different coolant flow rates (left: in-hole axis; right: inter-hole)}
\label{fig:temp_flowrate}
\end{figure}

\begin{figure}[!htbp]
\centering
\includegraphics[width=2.82152in,height=2.16535in]{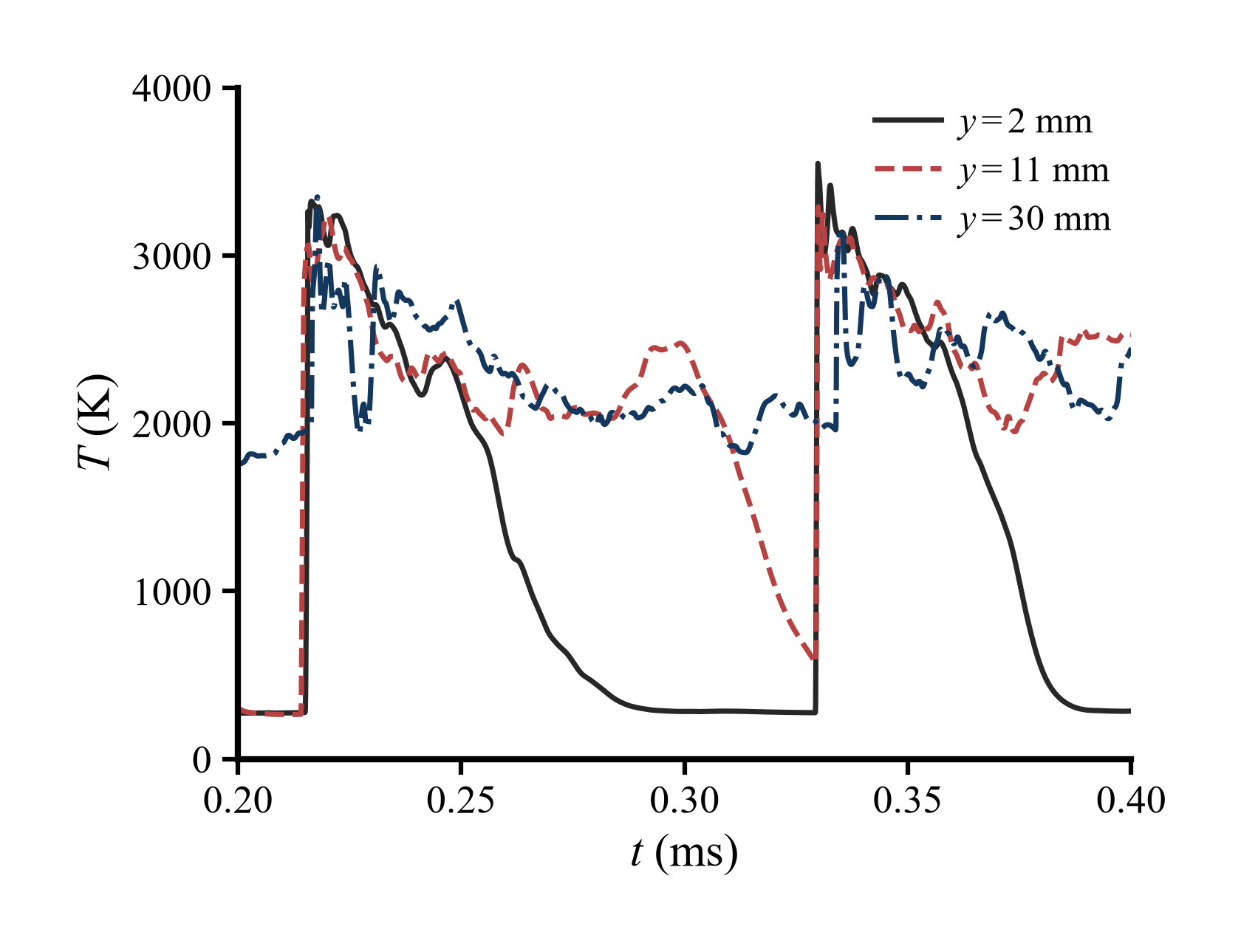}
\includegraphics[width=2.65468in,height=2.16535in]{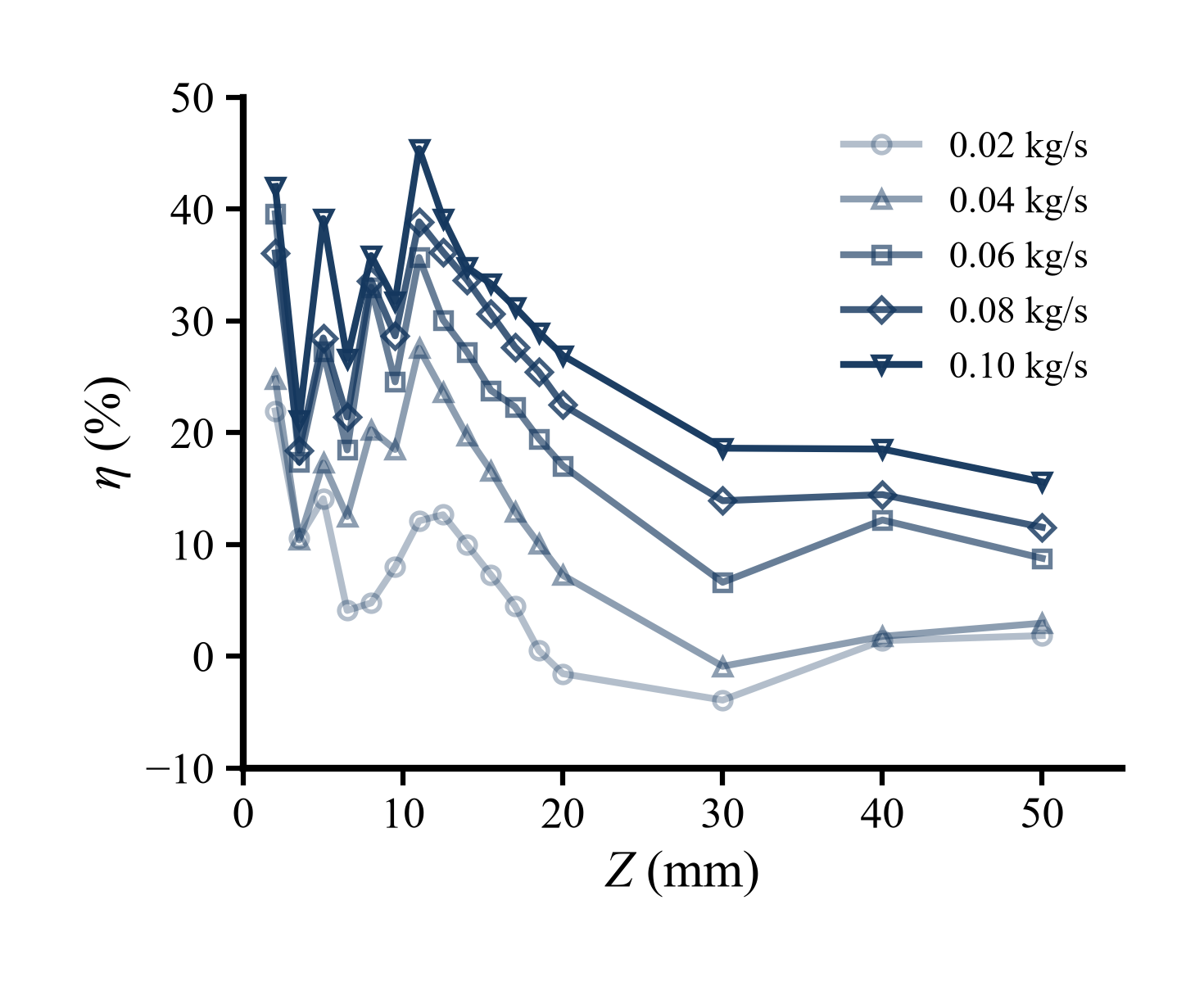}
\caption{(a) Wall temperature variation over one detonation cycle at different axial
heights, and (b) cooling effectiveness at different coolant flow rates}
\label{fig:cycle_temp}
\end{figure}

Figure~\ref{fig:temp_flowrate} presents the temperature distributions along the in-hole axis and the inter-hole axis. Within the region of $z = 0$--15~mm, the average temperature of the fluid near the wall
increases rapidly, while it remains at a relatively high level above 15~mm.
This height approximately corresponds to the detonation wave head height,
indicating that the fluid temperature is high downstream of the detonation
wave, and the thermal load on the wall increases significantly. This behavior is explained as follows. Upstream of the detonation wave, fresh mixture
exists at a low temperature (300~K), keeping the fluid temperature near the
wall in this region at a relatively low level. Within the range of heights
that the fresh mixture can reach, the lower the axial position, the longer the
fresh mixture occupies the near-wall region during each cycle, resulting in a
lower average temperature and larger temperature fluctuations. In contrast,
the region downstream of the detonation wave is filled with high-temperature
combustion products, which are further compressed and heated as they pass
through the oblique shock wave. Consequently, the average fluid temperature near the wall in this region remains high with smaller temperature fluctuations; Figure~\ref{fig:cycle_temp} shows the temperature variation over one cycle at different axial heights.

\begin{figure}[!htbp]
\centering
\includegraphics[width=5.76806in,height=3.35694in]{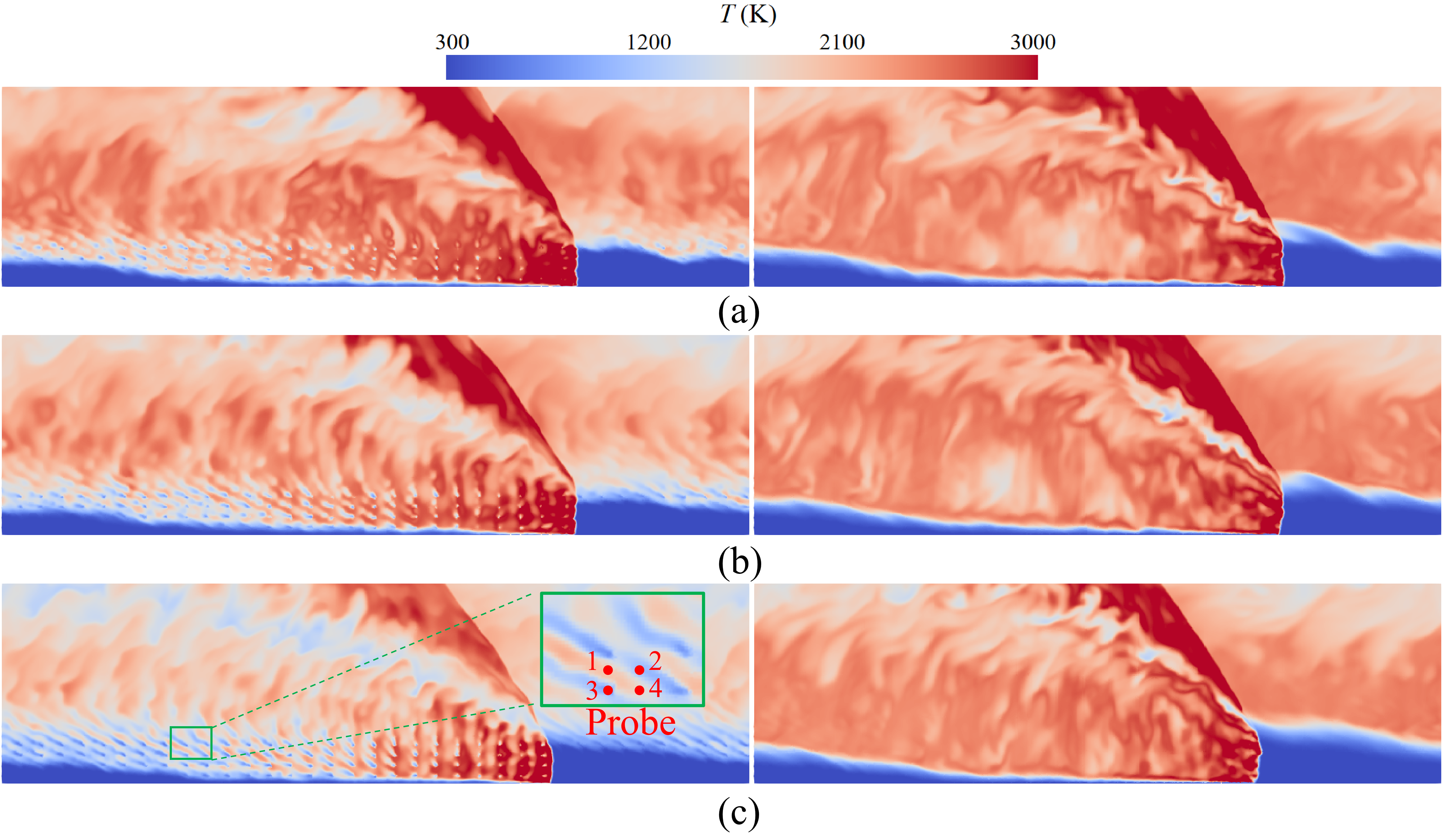}
\caption{Temperature contours at different coolant flow rates (left column: outer wall; right column: $R=33$~mm cross-section). Rows correspond to (a)~0.02~kg/s, (b)~0.04~kg/s, and (c)~0.08~kg/s}
\label{fig:contour_flowrate}
\end{figure}

The cooling effectiveness at different axial positions after the introduction of the coolant is shown in Figure~\ref{fig:cycle_temp}. In the upstream region of the combustor ($z\leq15$~mm) it is relatively high. However, once the coolant flow rate reaches 0.06~kg/s, a further increase no longer improves the cooling effectiveness significantly in the coolant-injection region ($z\leq15$~mm). Figure~\ref{fig:contour_flowrate} presents the unwrapped views of the near-wall fluid surface and of the cross-section at $R=33$~mm for different coolant flow rates, with four representative monitoring points marked.
Point~3 is located at the center of a cooling hole. Point~2 is at the center
of four adjacent cooling holes. Points~1 and~4 are located midway between two
axially adjacent holes and two circumferentially adjacent holes, respectively.
Based on the wall temperature contours and the axial fluctuations of the
average temperature, the following trends are evident. Point~3 exhibits
the most significant cooling effect. Point~2, located at the center of the
four holes, still experiences a noticeable cooling effect, as the coolant can
diffuse to this region under the influence of the flow field. Points~1 and~4,
located between adjacent holes, receive less coolant and thus exhibit
relatively weaker cooling effects. Consequently, in the axial temperature
distributions shown in Figure~\ref{fig:temp_flowrate}, a pronounced
sawtooth-like fluctuation appears in the region $z\leq15$~mm. As the coolant
flow rate increases, the sawtooth fluctuations in the wall temperature and the cooling effectiveness become more pronounced, and the temperature difference between the
hole locations and the mid-hole regions gradually increases. Furthermore, the temperature contours in Figure~\ref{fig:contour_flowrate} for the near-wall fluid surface (left) and the cross-section at $R=33$~mm (right) indicate that the coolant
significantly reduces the wall temperature, with the cooling effect becoming
more substantial at higher flow rates. In contrast, the temperature contours
at the cross-section of $R=33$~mm show no noticeable differences, indicating
that the wall-injected coolant produces a significant cooling effect near the
wall but has a relatively small impact on the internal flow field of the
combustor.

The coolant also provides some cooling in the downstream region ($z>15$~mm). As shown in Figure~\ref{fig:temp_flowrate},
once the coolant flow rate exceeds 0.04~kg/s, a significant cooling effect appears in the downstream wall region and becomes more pronounced as the flow rate increases. At a coolant flow rate of 0.10~kg/s,
the wall temperature decrease reaches approximately 400~K, and the cooling effectiveness reaches about 20\%, which demonstrates the feasibility of
applying liquid-fuel film cooling to RDEs.
However, at a coolant flow rate of 0.02~kg/s, no appreciable cooling effect is
observed in the downstream wall region relative to the no-cooling case; instead,
the wall temperature increases in localized regions, resulting in a negative cooling effectiveness. This occurs because at such a low flow rate the injected coolant does not form an effective insulating layer but instead acts as supplementary fuel in the downstream region, intensifying the local heat release.

\begin{figure}[!htbp]
\centering
\includegraphics[width=5.76806in,height=1.8625in]{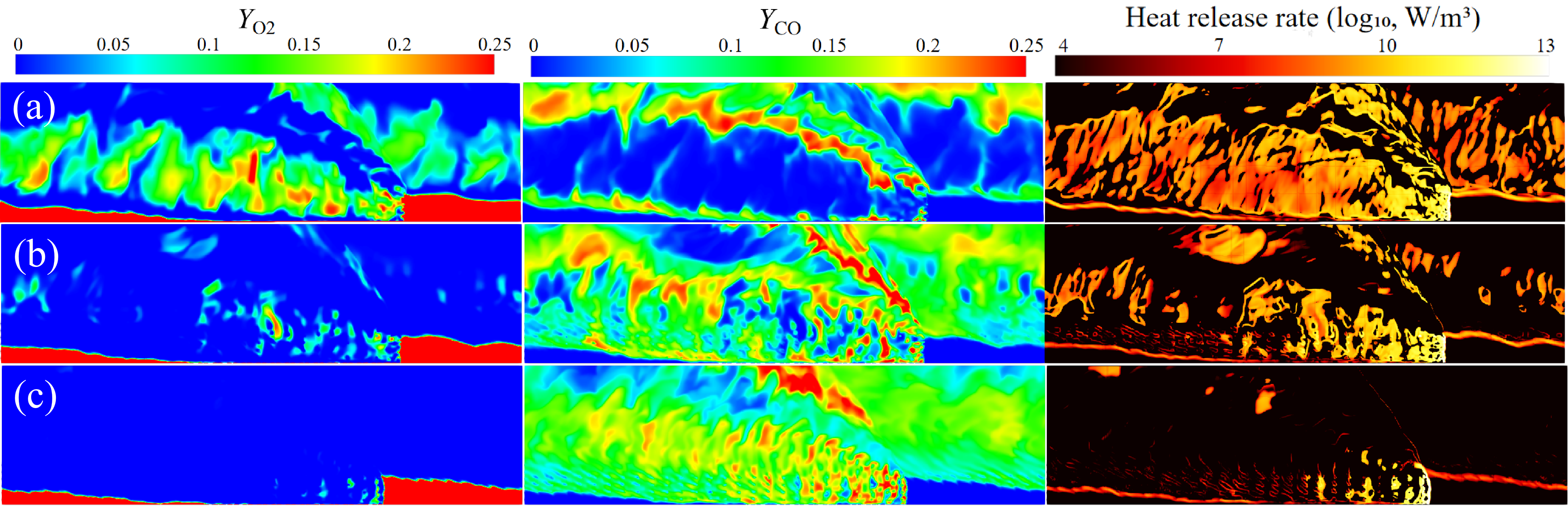}
\caption{O$_2$ and CO mass fractions and heat release rate contours at different coolant flow rates:
(a) no cooling, (b) 0.02~kg/s, and (c) 0.08~kg/s}
\label{fig:species_flowrate}
\end{figure}

Figure~\ref{fig:species_flowrate} presents the contours of O$_2$ and CO mass
fractions as well as the heat release rate at the near-wall fluid surface for different coolant flow rates. Under no-cooling conditions, the region behind the detonation wave
is oxygen-rich, and residual oxygen reacts with a small amount of CO in a
deflagration reaction, resulting in a high temperature near the wall. At a
coolant flow rate of 0.02~kg/s, the injected coolant increases the local fuel
supply. The coolant droplets absorb heat, vaporize, and then react with the residual oxygen behind the wave in a deflagration, failing to develop an effective insulating layer. Therefore, effective cooling is not
achieved, and the wall temperature even increases due to enhanced local
combustion. At a coolant flow rate of 0.08~kg/s, the injected coolant causes
the equivalence ratio near the wall to deviate significantly from the
stoichiometric value, and the oxygen is almost depleted after the detonation
wave passes, thereby substantially suppressing the deflagration reaction behind the
wave. Meanwhile, the continuously injected coolant evaporates and absorbs heat behind the wave while forming an insulating layer, resulting in a significantly
improved cooling effect.

\begin{figure}[!htbp]
\centering
\includegraphics[width=0.5\textwidth]{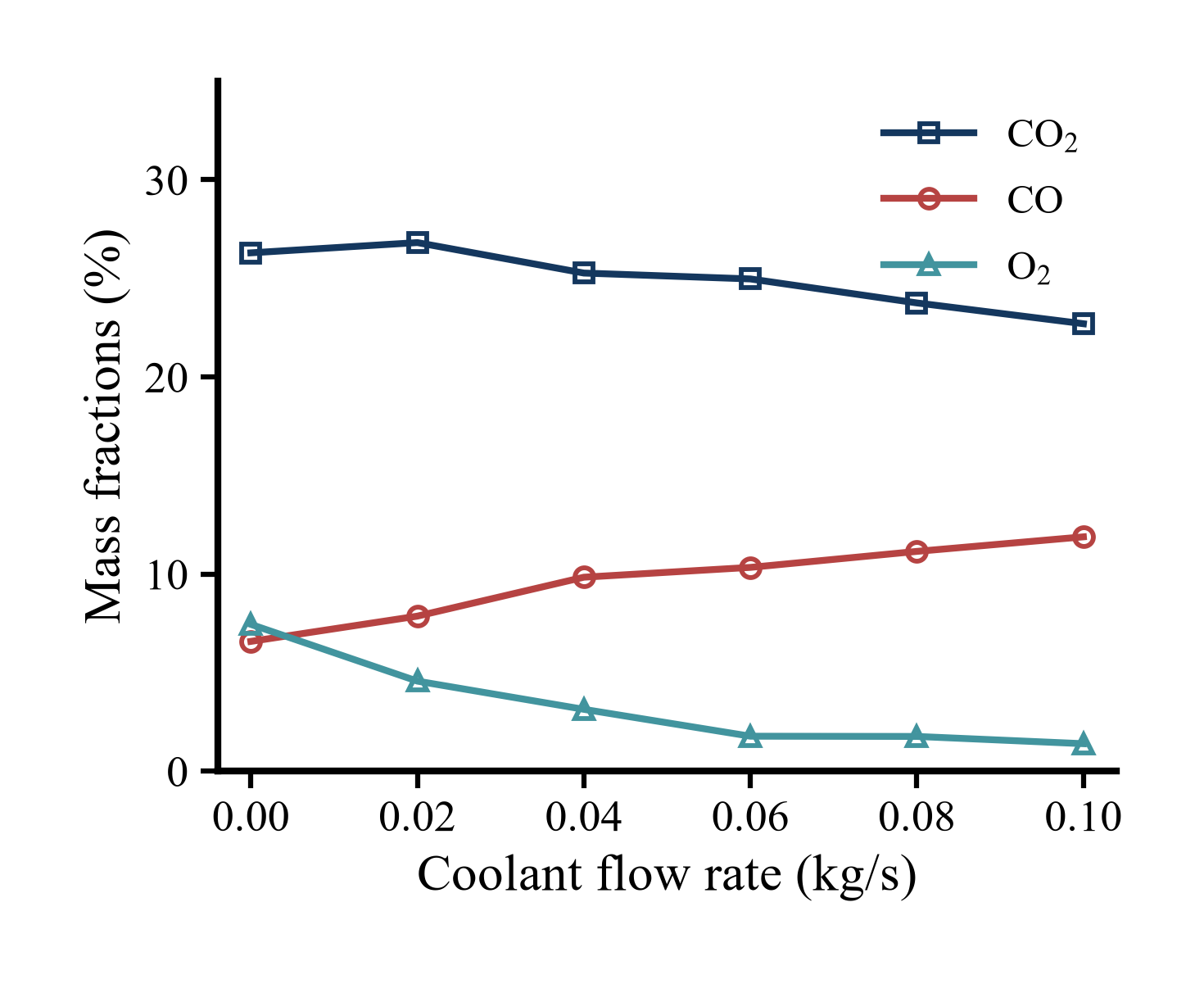}
\caption{Outlet species mass fractions as a function of coolant flow rate}
\label{fig:outlet_species}
\end{figure}

As the coolant flow rate increases, the overall equivalence ratio in the combustor rises, and the equivalence
ratio near the wall deviates significantly from the stoichiometric value,
leading to a decrease in combustion efficiency. To investigate the effect of the coolant flow rate on the combustion efficiency, the outlet mass fractions of the main species were monitored for different coolant flow rates. The variations in the mass fractions of oxygen, CO, and CO$_2$ are
shown in Figure~\ref{fig:outlet_species}. As the coolant flow rate increases,
the CO mass fraction at the outlet continuously increases, the O$_2$ mass
fraction continuously decreases, and the CO$_2$ mass fraction first increases
slightly at a coolant flow rate of 0.02~kg/s and then also decreases. This
trend can be explained as follows. A higher coolant flow rate increases the overall fuel input, which favors oxygen--fuel reactions and thus raises oxygen consumption.
However, the increased fuel supply also causes more oxygen to participate in
the first reaction step that produces CO, leaving relatively less oxygen
available for the further oxidation of CO to CO$_2$. Consequently, the outlet
species composition exhibits decreases in the O$_2$ and CO$_2$ mass fractions
and an increase in the CO mass fraction. In the kerosene mechanism adopted here, most of the heat is released in the second reaction step, in which CO is oxidized to CO$_2$. The incomplete combustion of fuel leads to a decrease in
combustion efficiency, which in turn causes a slight reduction in thrust.
Specifically, the combustor thrust is 1929.2~N without cooling, and decreases
to 1862.8~N at a coolant flow rate of 0.10~kg/s, representing a reduction of
3.44\%. This indicates that a higher fuel flow rate does not increase the thrust but instead reduces it. Therefore, a trade-off between the cooling effectiveness and the combustion efficiency is necessary in practical applications.

In summary, the coolant flow rate has a significant influence on both the
cooling performance and combustion characteristics of the RDE. An appropriate
increase in the coolant flow rate can effectively reduce the wall temperature
and form an insulating layer, thereby demonstrating the feasibility of
liquid-fuel film cooling in RDEs. However, an excessively low coolant flow
rate fails to provide effective cooling and instead intensifies local
combustion by supplying additional fuel, leading to an increase in wall
temperature. Meanwhile, although a further increase in the coolant flow rate
can enhance the cooling effect, it also results in decreased combustion
efficiency and a slight loss in thrust. Therefore, in practical engineering
applications, the coolant flow rate should be carefully selected based on
specific thermal protection requirements and propulsion performance demands
to achieve an optimal balance between the cooling effectiveness and the combustion efficiency.

\FloatBarrier
\subsection{Effect of atomization quality}

Liquid film cooling using liquid hydrocarbon fuels relies primarily on the
combined effects of heat absorption through vaporization and thermal isolation.
In the present simulations, the liquid hydrocarbon fuel enters the flow field as droplets. The droplets reduce the flow-field temperature both through the temperature difference between the droplets and the surrounding flow and through their latent heat of vaporization.
Meanwhile, the fuel vapor generated by evaporation forms a film near the wall,
which insulates the wall from the high-temperature products in the flow field.
It is therefore desirable that the droplets and their vapor remain close to the wall for more effective cooling, and the atomization quality is one of the key parameters that govern their spatial distribution. Here, different atomization conditions are represented by coolant droplets of different diameters, which allows the effect of atomization quality on the cooling performance to be examined.

\begin{figure}[!htbp]
\centering
\includegraphics[width=5.51218in,height=2.16669in]{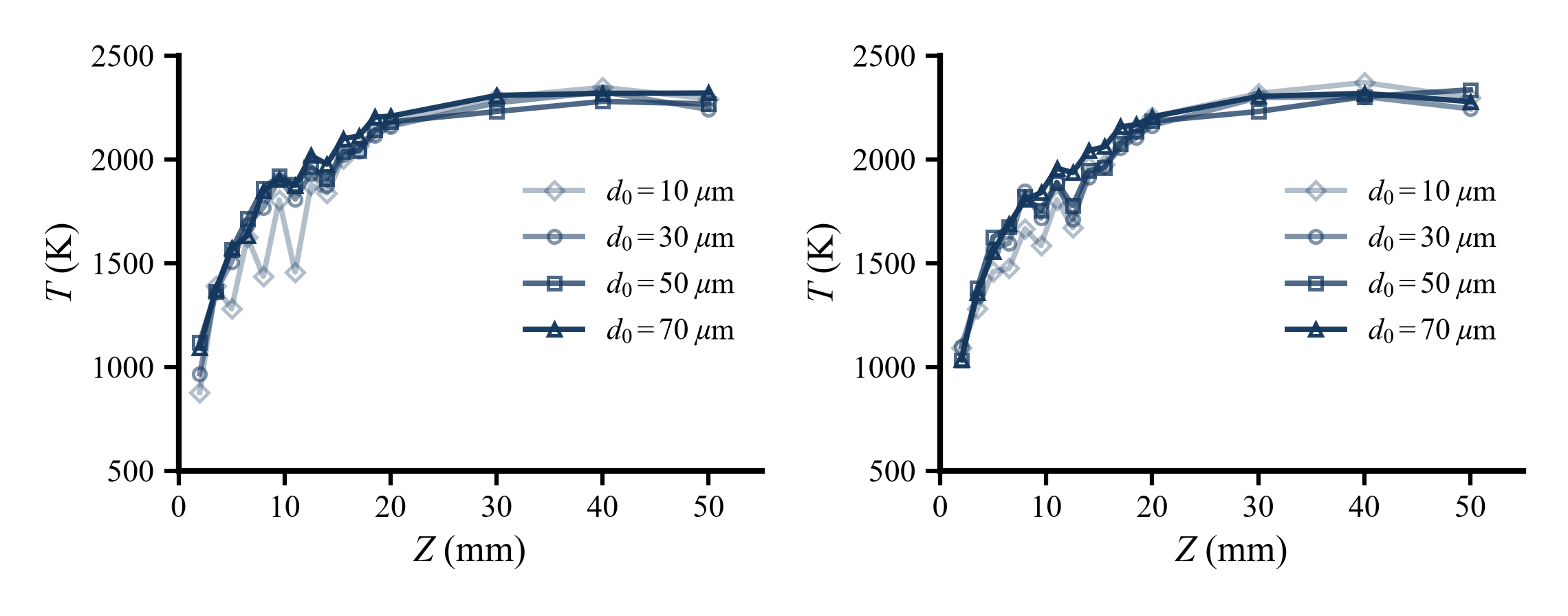}
\caption{Wall temperature distributions at different coolant droplet diameters (left: in-hole axis; right: inter-hole)}
\label{fig:temp_diameter}
\end{figure}

\begin{figure}[!htbp]
\centering
\includegraphics[width=2.6306in,height=2.16535in]{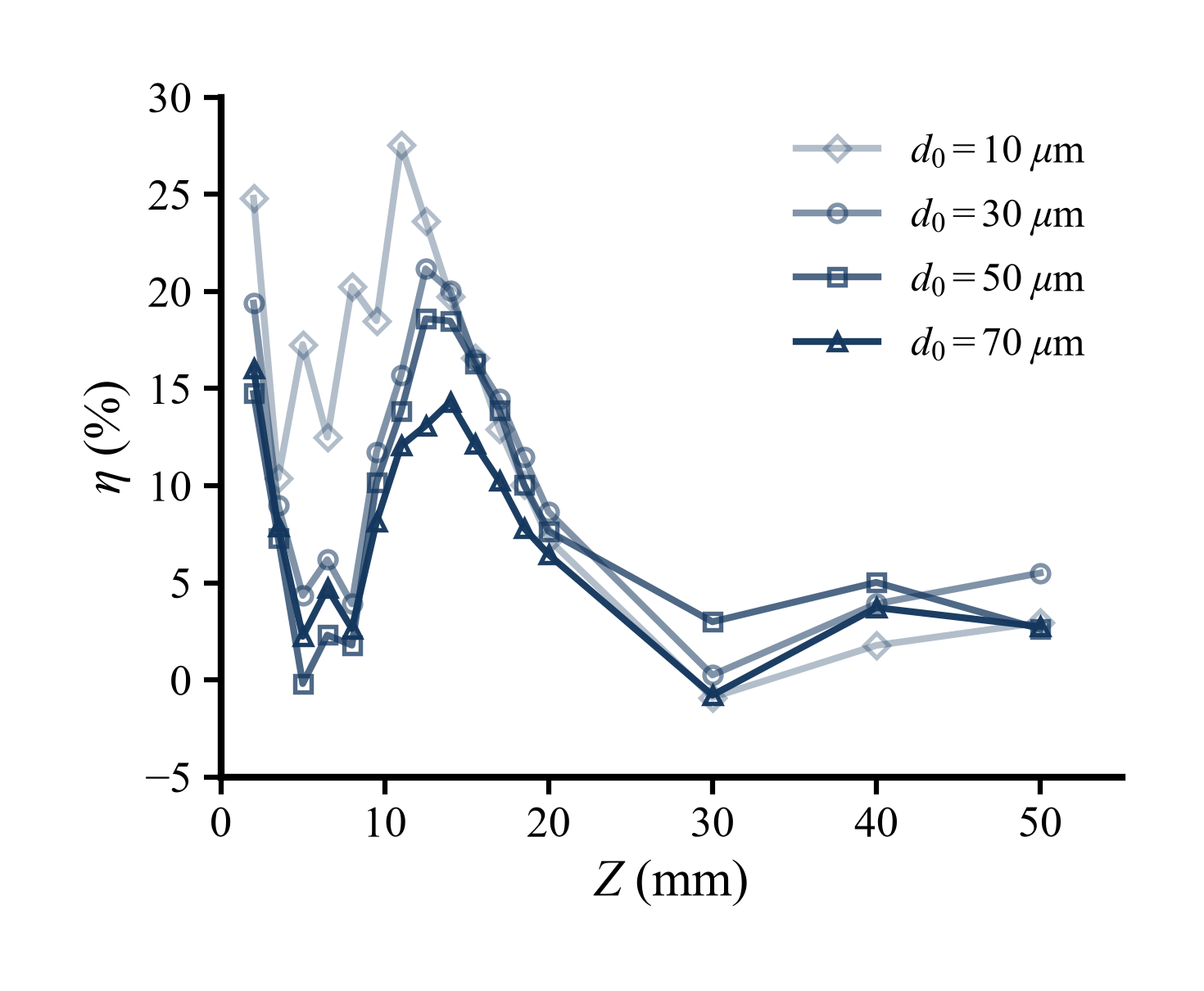}
\caption{Cooling effectiveness at different coolant droplet diameters}
\label{fig:cooling_diameter}
\end{figure}

\begin{figure}[!htbp]
\centering
\includegraphics[width=3.31326in,height=1.09737in]{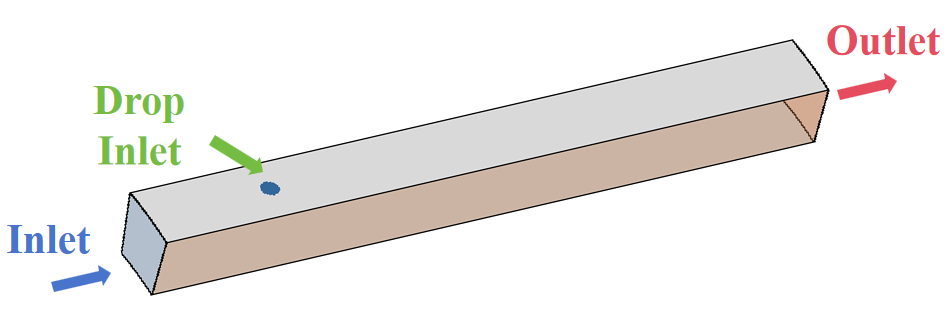}
\caption{Schematic of the 1/36 circumferential sector computational domain with a single cooling hole}
\label{fig:sector_domain}
\end{figure}

As shown in Figures~\ref{fig:temp_diameter} and~\ref{fig:cooling_diameter},
the droplet diameter has little effect on the cooling effectiveness in the downstream region of the combustor. Within the coolant-injection region, however, the 10~$\mu$m droplets give a significantly higher cooling effectiveness than droplets of other diameters. To elucidate the interaction mechanism
between droplet diameter and the flow field, a simplified model was established
by taking a 1/36 sector of the combustor in the circumferential direction, as
shown in Figure~\ref{fig:sector_domain}. In this simplified model, the inner
radius, outer radius, and axial length of the combustor remain unchanged, and
the cooling hole diameter is kept at 0.6~mm. Only one cooling hole is placed
at $z=11$~mm, with a coolant flow rate of 0.000278~kg/s. The droplets are injected perpendicularly into the flow field at 50~m/s, and the initial droplet diameter is the variable parameter.

To examine droplet behavior in the two environments considered here, the low-temperature fresh mixture region and the high-temperature post-detonation gas region, the periodic variation of the flow field is neglected and two sets of constant inlet and outlet conditions are used instead. For the low-temperature fresh
mixture region, the inlet conditions are set to match the combustor inlet
injection conditions: a total pressure of 8~atm, a total temperature of 300~K,
and an inlet gas composition of 0.5 O$_2$ + 0.5 N$_2$, with an outlet back
pressure of 6~atm. For the high-temperature post-detonation gas region, some simplification is required because the flow field expands rapidly and varies strongly. The inlet total pressure is set to
24~atm, the total temperature to 2500~K, and the inlet gas composition is
assumed to be 0.24 CO + 0.14 CO$_2$ + 0.41 N$_2$ + 0.21 H$_2$O, with oxygen
completely consumed, and the outlet back pressure is set to 20~atm.

\begin{figure}[!htbp]
\centering
\includegraphics[width=5.76806in,height=2.31528in]{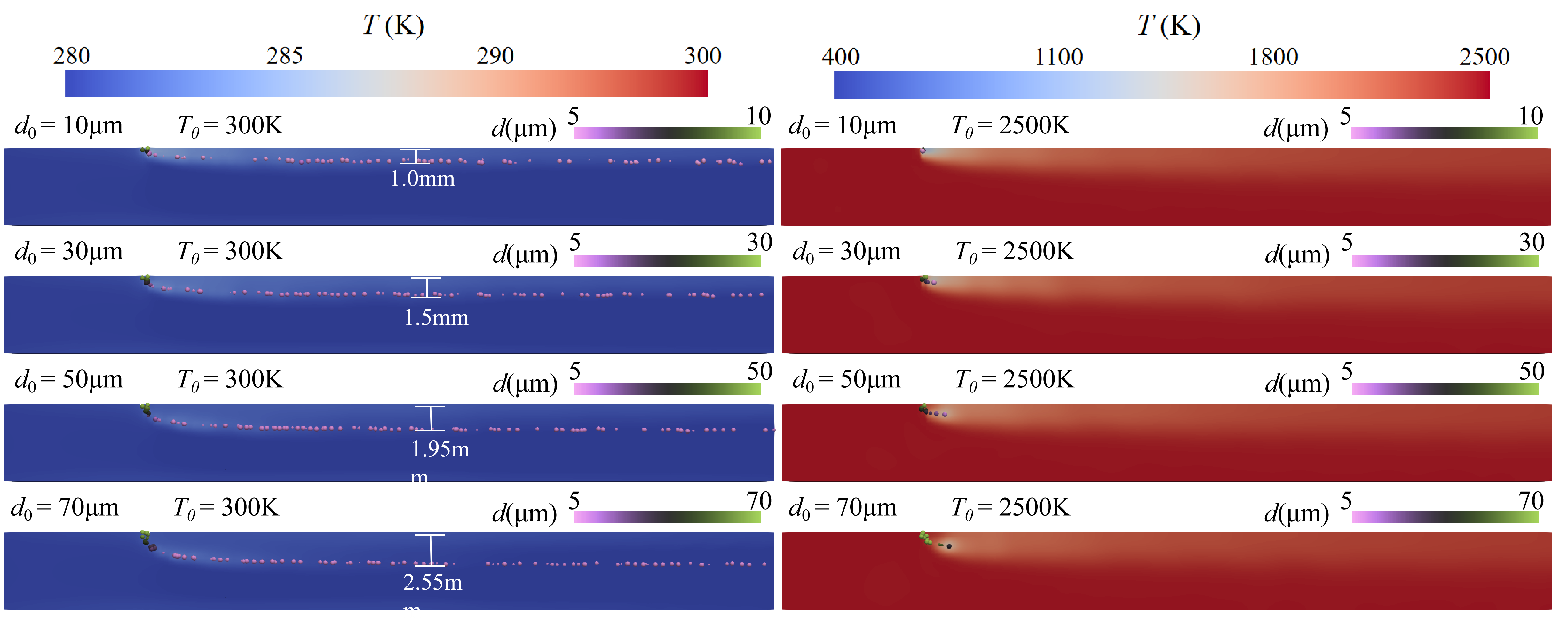}
\caption{Droplet distributions and temperature contours on the longitudinal section along the cooling hole axis:
(a) fresh mixture region and (b) high-temperature post-detonation gas region}
\label{fig:droplet_dist}
\end{figure}

The simulation results are shown in Figure~\ref{fig:droplet_dist}. The left
panel presents the results for the low-temperature fresh mixture region. For this case the total temperature of the incoming flow is 300~K, and at the high Mach number the static temperature is approximately 280~K, slightly below the droplet temperature. Consequently, the flow
field near the wall is actually heated by the coolant. After being injected
perpendicularly into the flow field, the droplets break up under the influence
of the flow and move downstream. The better the atomization quality (i.e., the
smaller the initial droplet diameter), the smaller the initial momentum of
each droplet and thus the smaller the penetration depth into the flow field.
For an initial droplet diameter of 70~$\mu$m, the penetration depth is
2.55~mm; for 10~$\mu$m droplets, the penetration depth is only 1~mm.
Small-diameter droplets remain clustered near the wall, so that when the
detonation wave passes, they can absorb heat and vaporize completely while
simultaneously forming an insulating layer that isolates the high-temperature
gas. This indicates that in the fresh mixture region, a higher degree of
atomization leads to better wall-attachment characteristics and thus better
cooling performance when the detonation wave sweeps over.

\begin{figure}[!htbp]
\centering
\includegraphics[width=3.02857in,height=2.46559in]{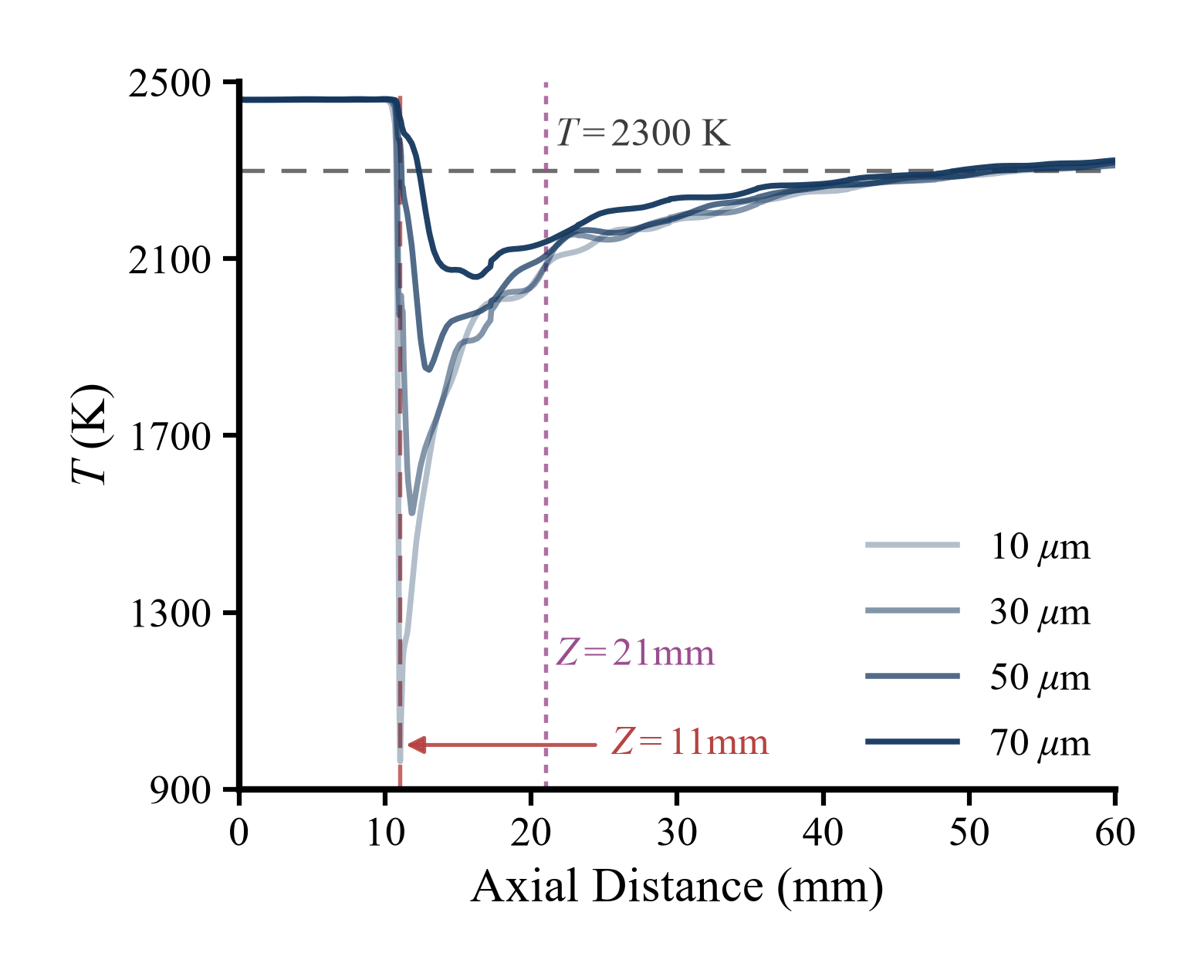}
\caption{Axial wall temperature distributions along the in-hole axis for different initial droplet diameters}
\label{fig:wall_temp_diameter}
\end{figure}

The right panel of Figure~\ref{fig:droplet_dist} presents the results for the
high-temperature post-detonation gas region. When the droplet diameter is small, the droplets evaporate rapidly at the high temperature, absorbing heat from the incoming flow and forming an insulating layer near the wall. As the droplet diameter increases, the
droplet lifetime becomes longer, and the penetration depth into the flow field
increases. This leads to a wider insulating layer, but the cooling gas becomes
more dispersed, reducing its ability to isolate the high-temperature gas. To evaluate the cooling effectiveness of droplets of different initial diameter more directly, Figure~\ref{fig:wall_temp_diameter} presents the axial wall temperature distribution for different initial diameters.
The droplets are injected into the flow field at $z=11$~mm, and therefore the
wall temperature decreases sharply at this location. The magnitude of the
temperature drop at the injection hole and the location of the minimum
temperature both vary with droplet diameter. For an initial diameter of
10~$\mu$m, the wall temperature at the injection hole ($z=11$~mm) reaches its
lowest value, with a temperature drop of approximately 1500~K. As the initial
diameter increases, the magnitude of the temperature drop gradually decreases,
and the location of the minimum temperature shifts downstream. For an initial
diameter of 70~$\mu$m, the maximum temperature drop is approximately 400~K,
and the minimum temperature occurs at $z=16$~mm. For all droplet diameters,
the same trend is observed: the wall temperature begins to decrease from the
cooling hole location, reaches a minimum, and then gradually recovers. At
$z=21$~mm, the wall temperature has already recovered to 2100~K, and the
differences in cooling effectiveness among different droplet diameters are no
longer significant. Near the outlet, the wall temperature recovers to 2300~K,
which is still approximately 160~K lower than the wall temperature upstream
of the cooling hole (2460~K). Thus, under the present flow-rate conditions, finer atomization gives a higher cooling effectiveness within about 10~mm downstream of the cooling hole, but the cooling effect weakens with distance and the differences between droplet diameters become negligible.

In summary, the atomization quality of the coolant is a key parameter affecting
the effectiveness of liquid film cooling. Under the conditions examined here, small droplets with an initial diameter of 10~$\mu$m give the best cooling performance within the coolant-injection region. This is attributable
to their low initial momentum and strong wall-attachment characteristics,
which enable them to rapidly vaporize and absorb heat while forming an
effective insulating layer when the detonation wave passes. As the droplet
diameter increases, the penetration depth increases, the cooling effectiveness
decreases, and the location of the minimum temperature shifts downstream. The
region of high cooling effectiveness is essentially limited to approximately
10~mm downstream of the cooling hole. Therefore, sufficiently fine atomization in engineering design improves the cooling effectiveness near the wall and enhances the reliability of the thermal protection.

\FloatBarrier
\subsection{Effect of fuel allocation between main flow and cooling}
The results presented in Section~\ref{sec:flowrate} demonstrate that when the
coolant flow rate exceeds 0.04~kg/s, further increasing the flow rate continues
to improve the overall cooling effectiveness across the combustor, although
the benefit within the injection region ($z\leq15$~mm) becomes less pronounced
beyond 0.06~kg/s. However, in practical engineering applications, the amount
of fuel available for the cooling system is limited. If too much fuel is
allocated to the cooling system, the main flow in the combustor may become
fuel-deficient, making it difficult to maintain a stable detonation. Therefore, the way in which the fuel is split between the main injection and the cooling holes is critical for balancing the propulsion and cooling performance. In this subsection, the
total fuel flow rate is kept constant at 0.17~kg/s. The flow rate allocated to
the cooling system is varied from 0 to 0.05~kg/s, with the corresponding main
flow rate varying from 0.17 to 0.12~kg/s. The effects of this fuel split on the cooling performance and the thrust of the combustor are then examined.
\begin{figure}[!htbp]
\centering
\includegraphics[width=5.44551in,height=2.16669in]{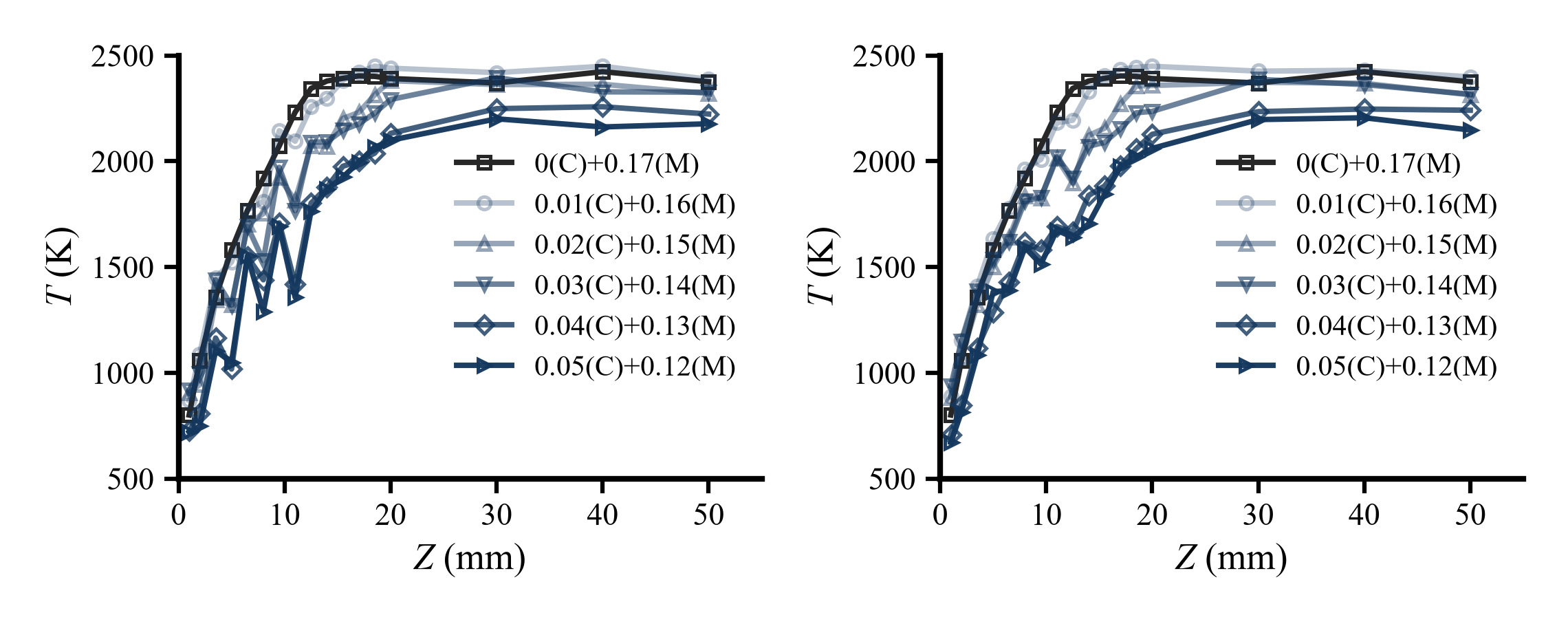}
\caption{Wall temperature distributions for different coolant allocation schemes (left: in-hole axis; right: inter-hole; C: coolant flow rate; M: main flow rate)}
\label{fig:temp_distribution}
\end{figure}

The axial distributions of wall temperature are shown in
Figure~\ref{fig:temp_distribution}. When the coolant flow rate exceeds
0.04~kg/s, a greater allocation of fuel to the cooling system leads to better
wall cooling effectiveness and a lower average wall temperature in the
combustor, which is consistent with the conclusions of
Section~\ref{sec:flowrate}. When the coolant flow rate is less than 0.04~kg/s,
the cooling effect is not significant. At coolant flow
rates of 0.02~kg/s and 0.03~kg/s, the average temperature in the upstream
region of the combustor is somewhat reduced, but the cooling effect in the
downstream region remains insignificant. At a coolant flow rate of 0.01~kg/s,
essentially no cooling effect is observed in the detonation wave region, and the
axial average wall temperature in the downstream region increases compared to
the no-cooling condition, with the overall average wall temperature also
rising. This further confirms the conclusion drawn in Section~\ref{sec:flowrate} that at low flow rates the heat absorbed by vaporization and the thermal isolation provided by the coolant are both weak, and the coolant may instead intensify the deflagration reactions near the wall and raise the wall temperature.

\begin{figure}[!htbp]
\centering
\includegraphics[width=0.5\textwidth]{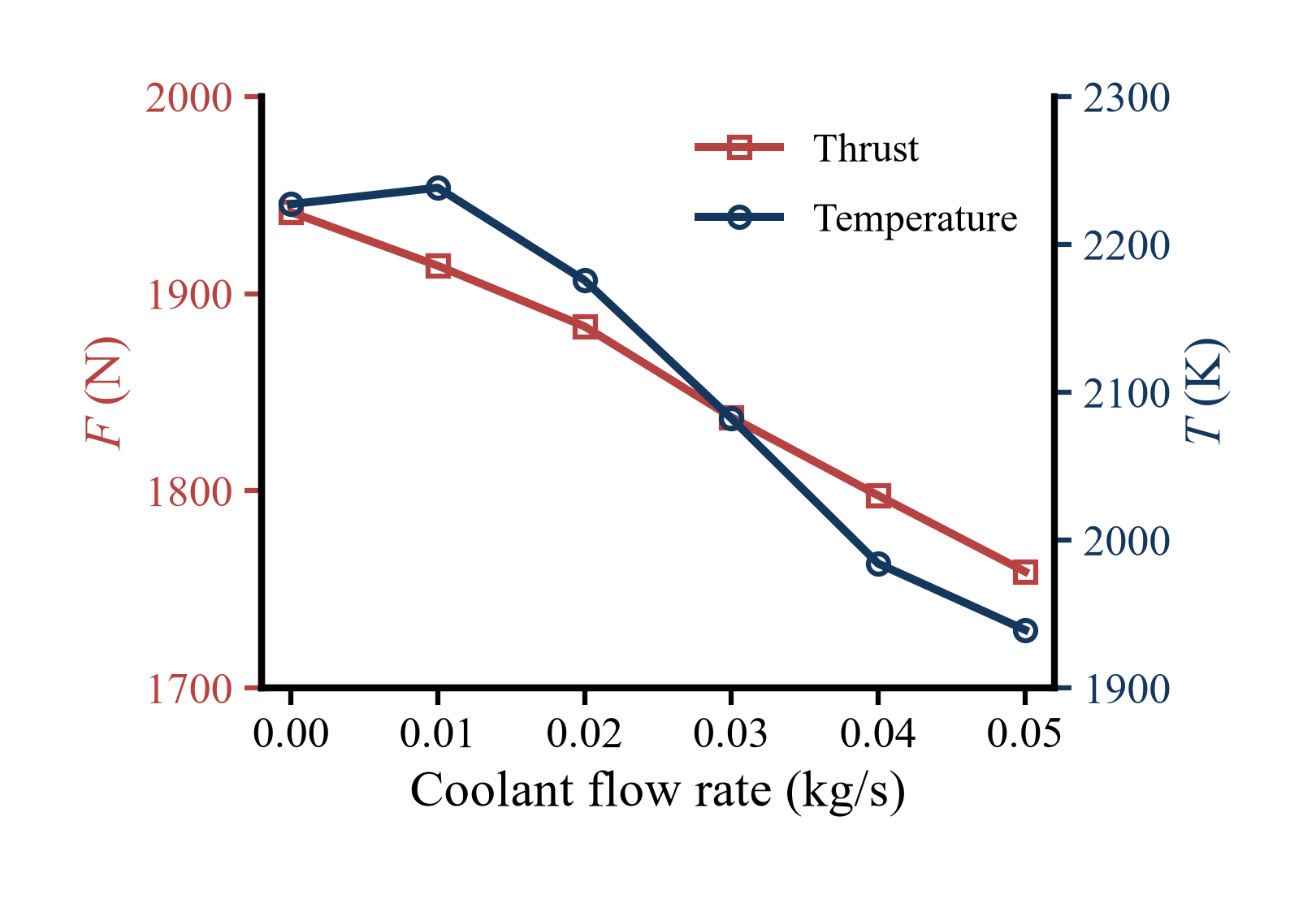}
\caption{Thrust and average wall temperature for different coolant allocation schemes}
\label{fig:thrust_distribution}
\end{figure}

In addition to the differences in cooling effectiveness, the thrust decreases as the coolant flow rate increases. Figure~\ref{fig:thrust_distribution} shows the thrust and the average wall temperature for different coolant flow rates. When all of the fuel is injected through the head with the main flow, the detonation velocity is 1828.2~m/s and the combustor thrust is 1941.3~N. At a coolant flow rate of 0.05~kg/s the detonation velocity decreases to 1702.6~m/s and the thrust to 1759.1~N, reductions of 6.9\% and 9.4\%, respectively, relative to the no-cooling case. Coolant injection thus reduces both the detonation velocity and the propulsion performance of the combustor. Therefore, in practical engineering
applications, cooling performance and propulsion performance must be
jointly considered to determine a reasonable coolant allocation.

In summary, the distribution of fuel between the main flow and the cooling
system has a significant impact on both the thermal protection performance and
propulsion performance of the combustor. When the fuel flow rate allocated to
the cooling system exceeds 0.04~kg/s, increasing the coolant flow rate
continues to reduce the overall average wall temperature and improve cooling
effectiveness, particularly in the downstream region, despite the saturation of the cooling effectiveness within the immediate injection zone. When the coolant flow
rate is too low ($\leq$~0.01~kg/s), the cooling effect is not significant,
and the wall temperature may even increase due to enhanced local combustion.
At the same time, an increase in the coolant flow rate inevitably reduces the detonation velocity and the combustor thrust. Therefore, in
practical engineering applications, a trade-off between cooling performance
and propulsion performance must be made based on specific mission requirements
to select the optimal fuel flow rate distribution scheme.

\section{Conclusions}
This paper reports a numerical study of wall cooling in a kerosene-fueled rotating detonation combustor with a head-end dual-purpose injection configuration, in which the film-cooling holes also serve as fuel injectors. The effects of coolant flow rate, atomization quality, and fuel allocation between the main combustor and the cooling system are systematically analyzed. The main conclusions are as follows:

(1) The coolant flow rate governs a competition between evaporative cooling and secondary combustion that is inherent to the dual-purpose configuration. When the coolant flow rate reaches 0.06~kg/s, a further increase brings only marginal additional cooling within the injection-dominated region ($z\leq15$~mm). In the downstream region ($z>15$~mm) the cooling effect continues to improve with increasing flow rate,
leading to a sustained reduction in the overall average wall temperature. An excessively low coolant flow rate of 0.02~kg/s fails to form an effective insulating layer and may even raise the wall temperature by intensifying the local deflagration reactions, because the injected kerosene acts as a supplementary fuel source.

(2) Atomization quality critically determines the cooling effectiveness of the
dual-purpose injection. Small droplets with a diameter of 10~$\mu$m, owing to
their low initial momentum and strong wall-attachment characteristics, rapidly vaporize and absorb heat while forming an effective insulating layer as the detonation wave passes; they therefore achieve a substantially higher cooling effectiveness than larger droplets. The region of high cooling effectiveness is essentially
limited to approximately 10~mm downstream of the injection holes.

(3) Under a fixed total fuel flow rate, the allocation of kerosene between the main injection and the dual-purpose cooling holes directly affects the propulsion performance. As the fuel flow rate allocated to the cooling system
increases, less fuel is available for the primary detonation combustion zone,
leading to a continuous decrease in both the detonation velocity and the combustor thrust. At a coolant flow rate of 0.05~kg/s, the thrust is reduced by 9.4\%
compared to the no-cooling condition. This trade-off is a unique characteristic of the head-end dual-purpose configuration, in which the coolant and the fuel are the same substance.

\bibliographystyle{unsrtnat}
\bibliography{references}

\end{document}